\documentclass[12pt]{elsarticle}

\usepackage[utf8]{inputenc}
\usepackage{times}
\usepackage{geometry}
\usepackage{tabularx}
\usepackage[titletoc]{appendix}
\usepackage{amsmath,amsfonts,amsthm,amssymb}
\usepackage{commath}
\usepackage{listings}
\usepackage{wrapfig}
\usepackage{graphicx}
\usepackage{amssymb}
\usepackage[pdftex,dvipsnames]{xcolor}
\usepackage{bm}
\usepackage{siunitx}
\usepackage{sidecap}
\usepackage{placeins}
\usepackage{mathtools}
\usepackage{lineno}
\usepackage{upgreek}
\usepackage[font={footnotesize}]{caption}

\newtheorem{definition}{Definition}

\RequirePackage{geometry}
\newcommand{\ignore}[1]{}  

\DeclareMathAlphabet{\mathpzc}{OT1}{pzc}{m}{it}

\setkeys{Gin}{width=0.9\linewidth,keepaspectratio}

\newcommand{\const}[1]{\mathrm{#1}}
\newcommand{\eqr}[1]{Eq.\thinspace(\ref{eq:#1})}

\newcommand{\eqrp}[1]{(Eq.\thinspace\ref{eq:#1})}
\newcommand{\fgr}[1]{Fig.\thinspace\ref{fig:#1}}
\newcommand{\pfrac}[2]{\frac{\partial #1}{\partial #2}}
\newcommand{\pfracb}[2]{\partial #1/\partial #2}

\usepackage{siunitx}
\usepackage{graphicx}
\usepackage{hyperref}
\hypersetup{
  colorlinks=true,
  citecolor=OliveGreen,
  linkcolor=RoyalBlue,
  urlcolor=NavyBlue,
  pdftitle = {Boundary Conditions for Continuum Simulations of Wall-bounded
  Kinetic Plasmas},
  pdfauthor = {Petr Cagas, Ammar Hakim, and Bhuvana Srinivasan}
}

\usepackage{tabularx}
\usepackage{mdwlist} 

\newcommand{\comment}[1]{\textit{\textcolor{red}{#1}}}
\renewcommand{\comment}[1]{}

\setkeys{Gin}{width=0.9\linewidth,keepaspectratio}

\journal{Journal of Computational Physics}

\begin{document}

\begin{frontmatter}

\title{Boundary Conditions for Continuum Simulations of
  Wall-bounded Kinetic Plasmas}


\author[vt]{Petr Cagas}
\author[pppl]{Ammar Hakim}
\author[vt]{Bhuvana Srinivasan}

\address[vt]{Virginia Tech, Blacksburg, VA, USA}
\address[pppl]{Princeton Plasma Physics Laboratory, Princeton, NJ, USA}

\begin{abstract}
Continuum kinetic simulations of plasmas, where the distribution
function of the species is directly discretized in phase-space,
permits fully kinetic simulations without the statistical noise of
particle-in-cell methods.  Recent advances in numerical algorithms
have made continuum kinetic simulations computationally competitive.
This work presents the first continuum kinetic description of
high-fidelity wall boundary conditions that utilize the readily
available particle distribution function.  The boundary condition is
realized through a reflection function that can capture a wide range of
cases from simple specular reflection to more involved first
principles models.  Examples with detailed discontinuous Galerkin
implementation are provided for secondary electron emission using
phenomenological and first-principles quantum-mechanical models.
Results presented in this work demonstrate the effect of secondary
electron emission on a classical plasma sheath.

\end{abstract}


\end{frontmatter}

\setcounter{secnumdepth}{1}
\setcounter{tocdepth}{1}

\section{Introduction}

Kinetic models of plasmas are necessary to capture processes that
occur at small spatial and temporal scales and depend on the shape of
a particle distribution.  An example of such a process is
collisionless Landau damping where an electromagnetic wave is damped
in a plasma resulting in a flattening of the particle distribution
around its phase velocity.  Kinetic simulations are most commonly
performed using particle-in-cell (PIC) methods \cite{Birdsall2004}.
However, continuum kinetic methods, which involve a direct
discretization of the particle distribution function in phase space,
are becoming more popular.  Continuum kinetic methods are not affected
by statistical noise and, with advances in numerical algorithms, are
becoming computationally competitive.

This work presents the first simulations of high-fidelity models for
electron emission (SEE) in wall-bounded plasmas using a continuum
kinetic method.  In wall-bounded plasmas, the formation and dynamics
of plasma sheaths is an important consideration to study
plasma-material interactions.  Plasma sheaths are narrow regions of
net space charge that occur where electrons and ions come into contact
with a solid surface.  The process of sheath formation results from
significant differences in electron and ion masses and, consequently,
their thermal flows.  The faster outflow of electrons gives rise to a
potential barrier, which equalizes electron and ion fluxes to the wall
\cite{Robertson2013}.  This behaviour can be reproduced in the
simplest case by setting the particle distribution function to zero
for both species, ions and electrons, at the edge of the domain
\cite{Cagas2017s,Cagas2018} representing an ideal sink.

Despite the small spatial scales associated with plasma sheaths, they
play an important role in particle momentum, energy, and heat transfer
and on surface erosion, which can have global effects on the plasma.
Furthermore, field-accelerated ions and hot electrons are known to
cause emission from the solid surface that can further alter the
system. One way to include SEE is a constant gain function
\cite{Campanell2012}; however, this technique does not account for the
dynamic role of the incoming distribution on the SEE. In reality, the
incident particles can be reflected back, penetrate the material and
then be rediffused with lower energy, or the electrons originally in
the material can gain energy from the incoming particles and be
released into the plasma.  Plasma sheaths and SEE can influence
material and plasma properties in any device where a surface contacts
a plasma such as in plasma thrusters \cite{Dunaevsky2003}, fusion
devices \cite{Takamura2004}, dielectric barrier and RF discharges
\cite{Mutsukura1990}, to name a few applications.

While there are many models addressing the SEE, they are often based
on complex coupling of different tools and/or a Monte-Carlo
technique. In this work, a generalized boundary condition
implementation for continuum kinetic methods is defined which directly
utilizes the information about the particle distribution functions and
enables straightforward implementation of various boundary
models. Examples using this boundary condition description are tested
on simplified boundary conditions and extended for high-fidelity
electron emission boundary conditions using a phenomenological model
\cite{Furman2002} as well as a first-principles based model
\cite{Bronold2015}.  The boundary conditions and infrastructure to
incorporate electron emission can be extended for other general
boundary conditions allowing for computationally efficient solutions
of physics-relevant surface models. The SEE boundary conditions and
results described in this work are presented using a discontinuous
Galerkin (DG) scheme that is extendable to arbitrarily high order,
however, the boundary condition descriptions are independent of the
numerical method.

The paper is organized as follows. Following the introduction, a brief
descriptions of plasma sheath physics and electron emission are
provided in Sec.~\ref{sec:wbp} and Sec.~\ref{sec:emission}
respectively.  Section~\ref{sec:bc} presents the description of a
continuum kinetic model and the general boundary conditions.  Specific
examples showing applications of a phenomenological and
first-principles models are in Sec.~\ref{sec:dg} with implementation
details for the discontinuous Galerkin continuum kinetic model in the
\texttt{Gkeyll} framework (\url{https://gkeyll.rtfd.org/}).  Results
are presented for the first continuum-kinetic plasma sheath
simulations using high-fidelity first-principles SEE boundary
conditions.

\section{Wall-bounded Plasmas \& Plasma Sheaths}\label{sec:wbp}

Interaction of plasma with a solid surface is typically governed by a
narrow region near the wall called a plasma sheath. Inside a sheath,
otherwise quasi-neutral plasma has a non-zero space charge.  The
charge is usually positive but inverse sheaths with negative charge
have been predicted in theory \cite{Campanell2016}.  The size of a
sheath is typically on the order of tens of Debye lengths with a Debye
length defined as, \footnote{In plasmas, electrons can usually move
  rapidly to shield any charge in the plasma.  The Debye length can be
  understood as a scale length of the shielded electric field
  exponential decrease with distance, $E(r) \propto
  \mathrm{exp}(-r/\lambda_{\const{D}})$.}
\begin{align*}
  \lambda_{\const{D}} = \sqrt{\frac{\varepsilon_0
      \const{k_B}T_{\const{e}}}{n_{\const{e}}q_{\const{e}}^2}},
\end{align*}
where $\varepsilon_0$ is vacuum permittivity, $T_{\const{e}}$ is
electron temperature in energetic units, $n_{\const{e}}$ is electron
number density, and $q_{\const{e}}$ is the electron charge.

Plasma sheaths form near a wall due to discrepancy in masses of plasma
species.  With comparable temperatures, this difference results in
different thermal velocities/fluxes.  Electrons, as the lightest
species in a plasma, are quickly absorbed into the wall
\citep{Robertson2013}.  The charge then gives rise to a potential
barrier, which works to equalize fluxes to the wall.  Despite the
microscopic nature of a sheath, it plays an important role in transfer
of particles, momentum, energy, and heat transfer and in surface
erosion, all of which can have global effects on the plasma.
Furthermore, field-accelerated ions and hot electrons are known to
cause an emission from the solid surface that can further alter the
system \cite{Mikellides2013}.

Formation of a sheath is strongly affected by boundary conditions at
the wall, such as electron emission, which is the main topic of this
paper.  For simplicity, an ideally absorbing wall is considered to
describe sheath and plasma behavior. For this simplified model, the
domain is divided into a quasi-neutral part where
$n_{\const{e}}=n_{\const{i}}=n_0$ and the non-neutral sheath with
monotonically decreasing potential, $\phi$, \cite{Langmuir1923,
  Riemann1990}.  The cold ions are assumed to enter the sheath region
with a non-zero velocity $u_{\const{i},0}$ and then "free-fall"
through the potential. The ion density inside the non-neutral region
is then obtained from the conservation of mass and
energy,\footnote{Conservation of mass and energy for ions inside the
  sheath,
\begin{align*}
  \underbracket{n_0u_{\const{i},0} =
    n_{\const{i}}(x)u_{\const{i}}(x)}_{\text{Conservation of mass}},
  \quad \underbracket{\frac{1}{2}m_{\const{i}}u_{\const{i},0}^2 =
    \frac{1}{2}m_{\const{i}}u_{\const{i}}(x)^2 +
    q_{\const{i}}\phi(x)}_{\text{Conservation of energy}},
\end{align*}}
\begin{align}\label{eq:ni}
  n_{\const{i}}(x) = n_0 \left( 1 -
  \frac{2q_{\const{i}}\phi(x)}{m_{\const{i}} u_{\const{i},0}^2}
  \right)^{-\frac{1}{2}}.
\end{align}
Electrons are assumed to instantly follow the electric potential,
\begin{align}\label{eq:ne}
  n_{\mathrm{e}}(x) = n_0
  \exp\left(-\frac{q_{\mathrm{e}}\phi}{\mathrm{k_B}T_{\mathrm{e}}}\right).
\end{align}
These densities are substituted into Poisson's equation,
\begin{align}\label{eq:poisson}
  \frac{\partial^2\phi(x)}{\partial x^2} =
  -\frac{n_{\mathrm{e}}(x)q_{\mathrm{e}} +
    n_{\mathrm{i}}(x)q_{\mathrm{i}}}{\varepsilon_0} =
  -\frac{n_0}{\varepsilon_0} \left[
    q_{\mathrm{e}}\exp\left(-\frac{q_{\mathrm{e}}\phi}{\mathrm{k_B}
      T_{\mathrm{e}}}\right) + q_{\mathrm{i}}\left( 1 -
    \frac{2q_{\mathrm{i}}\phi(x)}{m_{\mathrm{i}} u_{\mathrm{i},0}^2}
    \right)^{-\frac{1}{2}} \right].
\end{align}
Even though this equation cannot be solved analytically to obtain
sheath profiles, it leads to the classical Bohm sheath criterion
\citep{Bohm1949},
\begin{align}\label{eq:bohm}
  u_{\mathrm{i},0} \geq u_{\mathrm{B}} =
  \sqrt{\frac{Z\mathrm{k_B}T_{\mathrm{e}}}{m_{\mathrm{i}}}},
\end{align}
where $Z$ is ion ionization state. In this work, ions are assumed to
be singly ionized but have a non-zero temperature. Therefore, the Bohm
velocity used here has a slightly different form \cite{Tang2017},
\begin{align}\label{eq:bohm2}
  u_{\mathrm{i},0} \geq u_{\mathrm{B}} =
  \sqrt{\frac{\mathrm{k_B}T_{\mathrm{e}}+\upgamma
      \mathrm{k_B}T_{\mathrm{i}}}{m_{\mathrm{i}}}},
\end{align}
where $\upgamma$ is the heat capacity ratio, which can be defined
through the number of degrees of freedom, $N$, as $\upgamma =
(N+2)/N$.

Further physical insight into the Bohm criterion can be obtained from
\fgr{bohm}.  This figure presents a single electron density profile
(blue line) in the non-neutral sheath region together with three ion
densities based on $M=u_0/u_{\mathrm{B}}$.  If the ions are not
accelerated to the Bohm velocity in the presheath (red dashed line
denoting $M<1$), there is a point inside the sheath where the charge
density changes sign which contradicts the original assumption of a
monotonic potential drop.  This is further emphasized by the red
highlighted region in the plot. It is worth noting that this behavior
is significantly altered by presence of magnetic fields
\cite{Loizu2012} and by ionization inside the sheath and emission from
the wall \cite{Campanell2016}.
\begin{figure}[!htb]
  \centering
  \includegraphics[width=0.8\linewidth]{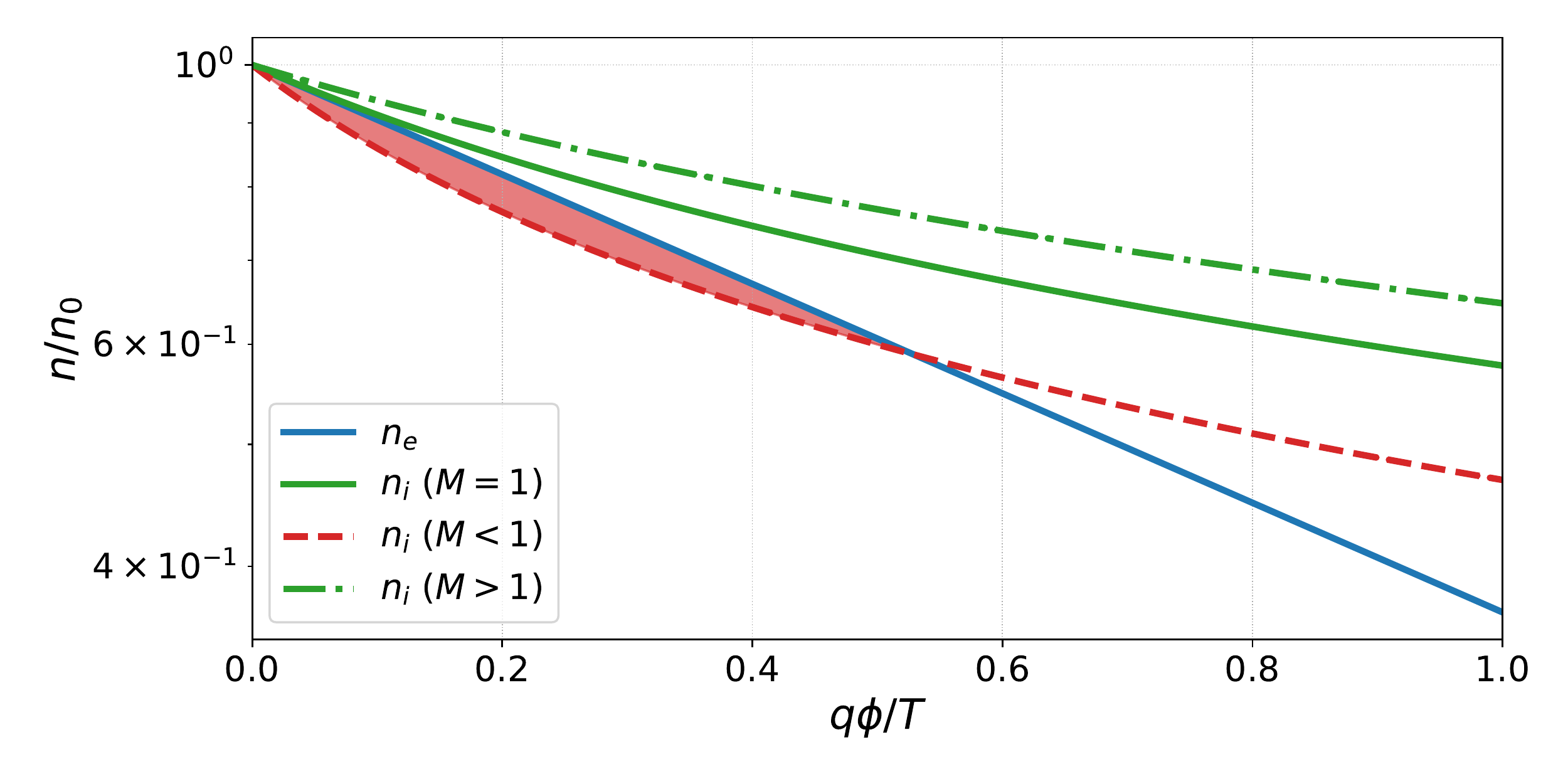}
  \caption{Electron (blue) and ion (green and red) densities inside
    the non-neutral sheath region as a function of sheath potential
    for different ion Mach numbers.  The ion densities are depicted
    based on the ratio of the ion velocity at the sheath entrance to
    the Bohm velocity. When the ions are not accelerated sufficiently
    in the presheath (red dashed line for $M<1$), the charge density
    changes sign inside the sheath, thus violating the original
    assumption as shown by the highlighted region. The figure is
    adapted from a Ph.D. dissertation \cite{Cagas2018}}
  \label{fig:bohm}
\end{figure}

A simplified model for sheath profiles including a uniform volumetric
source term, $S$, is described by Robertson
\cite{Robertson2013}. Solutions to this simplified model are presented
in \fgr{cls_robertson2} for normalized electron and ion densities,
electric field, potential, and ion bulk velocities. Ions, accelerated
by the presheath electric field, reach the Bohm velocity around
$8\,\lambda_{\const{D}}$ from the wall. The charge non-neutrality
becomes visually apparent around the same point as well.
\begin{figure}[!htb]
  \centering
  \includegraphics[width=0.8\linewidth]{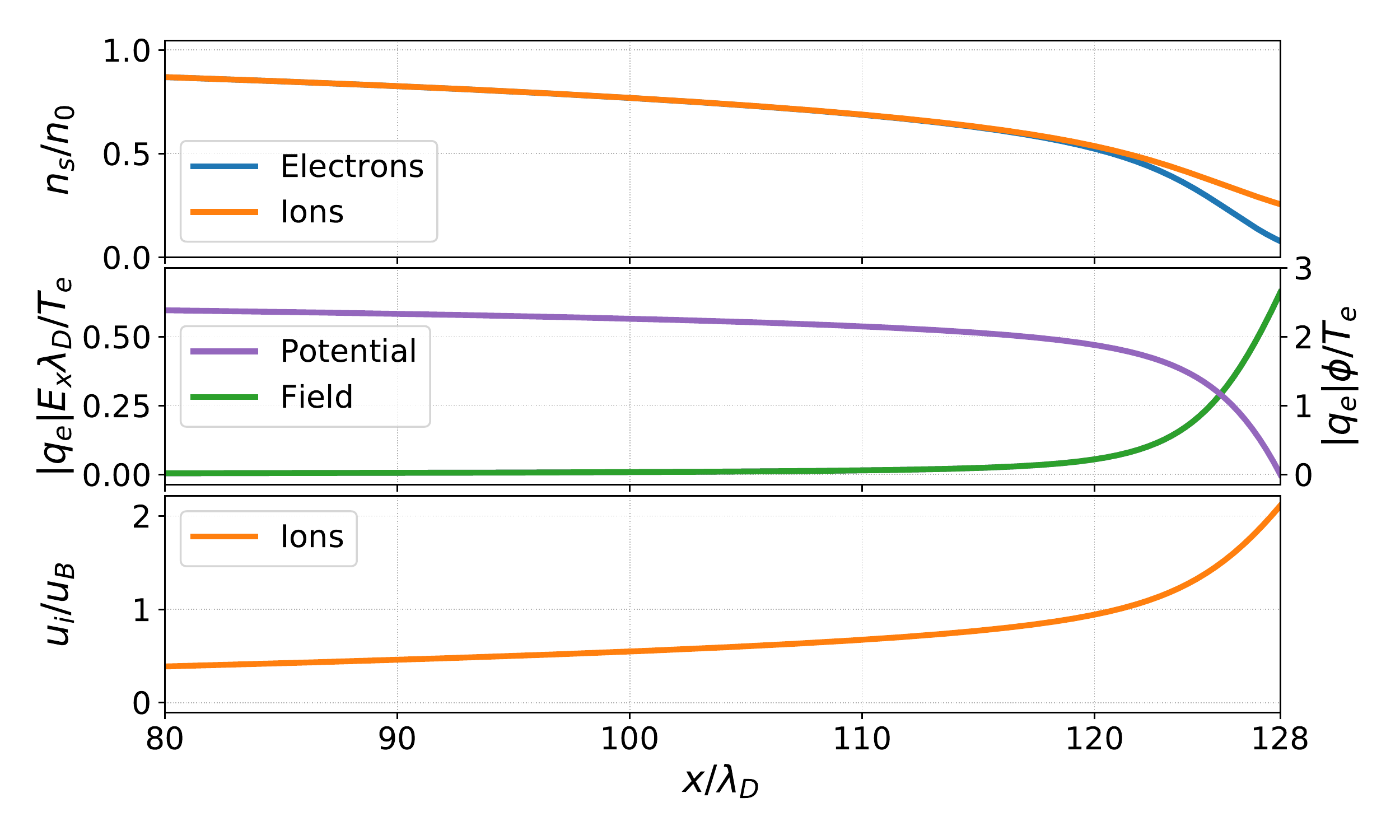}
  \caption{Simulated profiles of electron and ion densities, electric
    field and potential, and ion drift velocity obtained from a
    simplified sheath model by Robertson \cite{Robertson2013}. An
    ideally absorbing wall is located on the right side of the
    domain. Ions are accelerated by the presheath electric field and
    reach the Bohm velocity around $8\,\lambda_{\const{D}}$ from the
    wall. At the same point, the charge non-neutrality becomes
    apparent. The figure is adapted from a Ph.D. dissertation
    \cite{Cagas2018}.}
  \label{fig:cls_robertson2}
\end{figure}

\section{Electron Emission from a Wall}\label{sec:emission}

The treatment of a wall as an ideal absorber is often an unphysical
approximation.  In reality, some incoming particles are reflected from
the wall and some particles originating in the wall may enter the
plasma. The particles originating from the wall need to gain energy to
cross the surface potential barrier of the material.  One pathway is
by direct \cite{Furman2002} or indirect \cite{Bronold2018} energy
transfer from the incoming particles. Alternatively, the particles can
gain energy by wall heating or incoming electromagnetic radiation.

A common way to quantify the electron emission is through emission
yield, $\gamma$, which is defined as a ratio between the outgoing and
incoming fluxes.\footnote{Note that the emission yield, $\gamma$, is
  different from the heat capacity ratio, $\upgamma$, mentioned in
  Sec.~\ref{sec:emission}.  The later is not used in the rest of this
  work.}  Yields depend on material and incoming particle energy.  For
example, incoming particles with energy below \SI{600}{eV} have a
maximum yield of 0.56 for lithium, 0.97 for aluminum, and 1.27 for
copper \cite{Bruining1938}.

Particle emission from a wall can have significant effects on the
sheath and consequently on global plasma behavior.  With emission
yields above zero, $\gamma > 0$, the sheath can reach a space-charge
limitation \cite{Schwager1993,Stangeby2000,Sheehan2013}. For
$\gamma>1$, there are predictions that sheath potential can reverse
entirely \cite{Campanell2016}.

\section{General Boundary Conditions for Kinetic Plasmas}\label{sec:bc}

The models presented in Section~\ref{sec:wbp} do not properly account
for velocity distributions of particles.  When the plasma satisfies a
Maxwellian distribution, fluid models are often sufficient to describe
the dynamics.  However, the distribution inside a sheath is
non-Maxwellian \cite{Sydorenko2006, Kaganovich2007} and fully kinetic
models are needed.

A fully kinetic model can be derived from a continuous description of
discrete particles,
\begin{align}\label{eq:density}
  N_s(t,\bm{x},\bm{v}) = \sum_{i} 
  \delta\big(\bm{x}-\bm{X}_i(t)\big)
  \delta\big(\bm{v}-\bm{V}_i(t)\big),
\end{align}
where the sum is performed over all the particles of the same species
$s$.  The vectors $\bm{X}_i$ and $\bm{V}_i$ are positions and
velocities for all the particles $i$.  Taking a time derivative of $N$
and substituting the definition of velocity $\bm{\dot{X}}_i(t) =
\bm{V}_i(t)$ and Newton's second law with the Lorentz force,
\begin{align}\label{eq:motion}
  m_s\bm{\dot{V}}_i =
  q_s\bm{E}^m\big(t, \bm{X}_i(t)\big) + q_s \bm{V}_i(t)
  \times \bm{B}^m\big(t,\bm{X}_i(t)\big),
\end{align}
leads to the Klimontovich equation \cite{Nicholson1983},
\begin{align}\label{eq:klimontovich}
  \pfrac{N_s(t,\bm{x},\bm{v})}{t} +
  \bm{v}\cdot\nabla_{\bm{x}} N_s +
  \frac{q_s}{m_s}\left(\bm{E}^m + \bm{v} \times
  \bm{B}^m\right) \cdot\nabla_{\bm{v}} N_s = 0.
\end{align}
Knowing the electromagnetic field, \eqr{motion} fully describes a
collection of particles in continuous phase space. However, the
density, $N$, is still a sum of Dirac $\delta$-functions.  It can be
approximated by a smooth function by taking an ensemble average,
$f_s(\bm{x},\bm{v},t) := \left\langle N_s(\bm{x},\bm{v},t)
\right\rangle$, \cite{Nicholson1983}. $f$ is referred to as the
particle distribution function. The Klimontovich equation
\eqrp{klimontovich} then transforms into the Boltzmann equation,
\begin{align}\label{eq:boltzmann}
  \pfrac{f_s}{t} +
  \bm{v}\cdot\nabla_{\bm{x}}f_s +
  \frac{q_s}{m_s}\left(\bm{E} +
  \bm{v}\times\bm{B}\right)\cdot\nabla_{\bm{v}}f_s = \underbracket{
  -\frac{q_s}{m_s}\big\langle\left(\delta\bm{E} +
  \bm{v}\times\delta\bm{B}\right)\cdot\nabla_{\bm{v}}\delta
  N_s\big\rangle}_{\sum\left(\pfrac{f_s}{t}\right)_c},
\end{align}
where $\delta N_s(\bm{x},\bm{v},t) = N_s(\bm{x},\bm{v},t)-
f_s(\bm{x},\bm{v},t)$, etc. The term on the right-hand-side of
\eqr{boltzmann} corresponds to the intrinsically discrete nature of
particles like collisions \cite{Rosenbluth1957}. When the RHS is zero,
the equation is referred to as the Vlasov equation.  Note that the
individual species, $s$, are evolved separately. They are coupled with
electrostatic or electromagnetic fields and aforementioned collisions.

\fgr{sheath_distf} presents the distribution functions for ions and
electrons constructed from the density and momentum in
\fgr{cls_robertson2} assuming a Maxwellian distribution with constant
temperature across the domain.\footnote{Note that, as mentioned
  previously, the assumption of a Maxwellian distribution is violated
  inside a sheath and temperature typically drops due to decompression
  cooling \cite{Tang2011}. Nevertheless, these distribution functions
  can be used as a good approximation for an initial condition of a
  simulation.  In comparison to initialization with uniform
  conditions, this approach significantly limits excitation of
  artificial Langmuir waves which are otherwise present
  \cite{Cagas2018}.} The green line in the bottom panel marks the bulk
velocity of the ions and the dashed white line represents the Bohm
speed. Note that the point where the ion bulk velocity reaches the
Bohm speed is consistent with \fgr{cls_robertson2}.
\begin{figure}[!htb]
  \centering
  \includegraphics[width=0.8\linewidth]{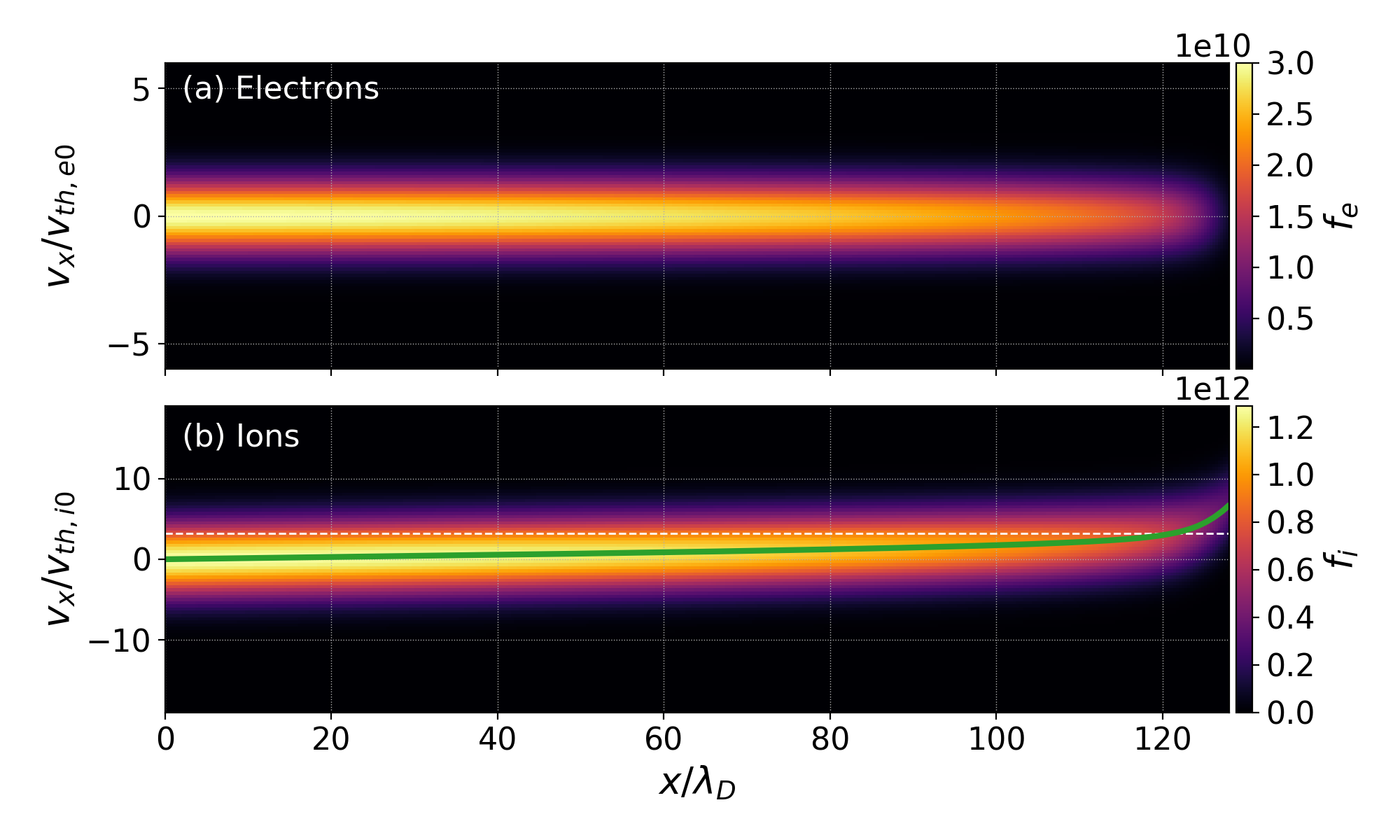}
  \caption{Electron (top) and ion (bottom) distribution functions
    constructed from the density and momentum presented in
    \fgr{cls_robertson2}. An absorbing wall is on the right side of
    the domain.  The green line in the bottom panel marks the bulk
    velocity of the ions and the dashed white line represents the Bohm
    speed.}
  \label{fig:sheath_distf}
\end{figure}

As shown in previous work \cite{Cagas2017s}, classical sheath results
using ideally absorbing walls as boundaries (mentioned in
Section~\ref{sec:wbp}) can be reproduced by setting the outgoing
distribution to zero at the wall, $f_{\const{out}}(t,\,
\bm{x}=\bm{x}_{\const{wall}},\, \bm{v})=0$.  These results also
confirm, that the electron distribution function near the wall is not
Maxwellian.  Furthermore, the availability of full particle
distribution functions, which are not contaminated by statistical
noise, near a wall can be used for more complex boundary conditions
when the particles are not simply absorbed.

Following is a general formulation of a boundary condition for a
distribution function near a wall assuming the entire incoming
(incoming implies to the wall) distribution function is known.  The
definition also includes a source term for effects like thermionic
emission which are not directly functions of the incoming distribution
but rather parameters of the wall, e.g., its temperature.  These terms
will be addressed in future work.

\begin{definition}
  Distribution function of particles coming out of a wall,
  $f_{\const{out}}$, is given by the incoming distribution function,
  $f_{\const{in}}$ and a reflection function, $R$,
  \begin{equation}\label{eq:genbc}
    f_{\const{out}}(t,\, \bm{x}=\bm{x}_{\const{wall}},\, \bm{v}) =
    \iiint_{\mathcal{V}_{\const{in}}} R(\bm{v},\,\bm{v}')
    f_{\const{in}}(t,\, \bm{x}'=\bm{x}_{\const{wall}},\,\bm{v}')
    \,\dif\bm{v}' + f_{\const{source}}(T_{\const{wall}},...), \quad
    \forall \bm{v} \in \mathcal{V}_{\const{out}}
  \end{equation}
  where the integration is performed over the velocities coming to the
  wall, $\bm{v} \in \mathcal{V}_{\const{in}}$. $f_{\const{source}}$ is
  a source term including effects like thermionic emission, which are
  generally functions of the wall conditions.
\end{definition}

Usage of this boundary condition can be demonstrated on specular
reflection.  Without loss of generality, the wall is assumed
perpendicular to the $x$-direction.  The reflection function is then
defined using Dirac $\delta$-functions,
\begin{equation}\label{eq:specular}
  R_{\const{x}}(\bm{v},\bm{v}') = \delta(v_{\const{x}}+v_{\const{x}}')
  \delta(v_{\const{y}}-v_{\const{y}}')\delta(v_{\const{z}}-v_{\const{z}}').
\end{equation}
For this reflection function, the integral \eqr{genbc} can be
calculated analytically and the boundary condition gives the expected
result,
\begin{equation*}\begin{aligned}
  f_{\text{out}}(v_{\const{x}},v_{\const{y}},v_{\const{z}}) &=
  \iiint_{\mathcal{V}_{\text{in}}}
  \delta(v_{\const{x}}+v_{\const{x}}')
  \delta(v_{\const{y}}-v_{\const{y}}')\delta(v_{\const{z}}-v_{\const{z}}')
  f_{\text{in}}(v_{\const{x}}',v_{\const{y}}',v_{\const{z}}') \,\dif
  v_{\const{x}}'\dif v_{\const{y}}'\dif v_{\const{z}}',\\ &=
  f_{\text{in}}(-v_{\const{x}},v_{\const{y}},v_{\const{z}}).
\end{aligned}\end{equation*}

This example is included here to demonstrate the approach. The
reflection function, $R$, can be replaced by more complex models using
the same framework. Presented below are two special cases of a
reflection function for electron emission using a phenomenological
model and a first-principles quantum mechanical model.

\subsection{Electron Emission Boundary Condition: Furman \& Pivi (2002) model}

Furman \& Pivi \cite{Furman2002} describe a widely used and referenced
phenomenological model, which uses analytical descriptions for three
populations of electron emission -- elastically reflected electrons,
rediffused electrons, and true-secondary electrons.  These species are
assumed to be produced by a mono-energetic (cold) beam of incoming
electrons.  For each incident beam with current $I_{\text{in}}$, the
model defines energetic distribution of electron yield, $\gamma =
I_{\text{out}}/I_{\text{in}}$,
\begin{equation}
  \pfrac{\gamma}{E} = \pfrac{\gamma_{\const{e}}}{E} +
  \pfrac{\gamma_{\const{r}}}{E} + \pfrac{\gamma_{\const{ts}}}{E},
\end{equation}
where $\gamma_{\const{e}}$, $\gamma_{\const{r}}$, and
$\gamma_{\const{ts}}$ correspond to the three aforementioned
populations.  An analytical profile for each population is determined
based on underlining physical properties and experimental data.

The first described group consists of primary electrons
semi-elastically reflected from the material surface.  Since they are
assumed not to lose any energy or only a small amount, the model
approximates this population with a narrow half-Gaussian centered
around the incoming energy.  Note that since the secondary electrons
cannot have higher energy than the incident ones (unless additional
energy is provided, for example, by heating), the distribution is
cropped at the incoming energy.  The contribution of the reflected
electrons is given as,
\begin{gather}\label{eq:furman_e}
  \pfrac{\gamma_{\const{e}}}{E}(E,E'\mu') =
  \theta(E)\theta(E'-E)\,\gamma_{\const{e}0}(E') \left[1 +
    e_1\left(1-\mu'^{e_2}\right)\right]
  \frac{2\exp\big(-(E-E')^2/2\sigma_{\const{e}}^2\big)}{\sqrt{2\pi}
    \sigma_{\const{e}}\mathrm{erf}\left(E'/\sqrt{2}\sigma_{\const{e}}\right)},
  \\ \gamma_{\const{e}0}(E') = P_{1,\const{e}}(\infty) + \left[\hat{P}_{1,\const{e}} -
    P_{1,\const{e}}(\infty)\right]
  \exp\left[\left(|E'-\hat{E}_{\const{e}}|/W\right)^p/p\right],\nonumber
\end{gather}
where $\theta(\cdot)$ is the Heaviside step function ensuring that the
incoming energy is higher than the outgoing.  $e_1$, $e_2$,
$\sigma_e$, $P_{1,\const{e}}(\infty)$, $\hat{P}_{1,\const{e}}$, $W$,
$\hat{E}_{\const{e}}$, and $p$ are fitting parameters.  $\mu$ and
$\mu'$ are direction cosines for the outgoing and incoming angles,
respectively.

The rest of the incident electrons are assumed to penetrate the
material.  As they interact with the material, they lose energy.  A
part of them eventually penetrate through the material potential
barrier again and return to the plasma.  These rediffused electrons
can have a wide range of energies between zero and the incident
energy,
\begin{gather}\label{eq:furman_r}
  \pfrac{\gamma_{\const{r}}}{E}(E,E,\mu') = \theta(E)\theta(E'-E)
  \gamma_{\const{r}0}(E') \left[1 + r_1\left(1-\mu'^{r_2}\right)\right]
  \frac{(q+1)E^q}{E'^{q+1}}, \\ \gamma_{\const{r}0}(E') = P_{1,\const{r}}(\infty)
  \left[1 - \exp\big(-(E'/E_{\const{r}})^r\big) \right], \nonumber
\end{gather}
where $r$, $r_1$, $r_2$, $q$, $P_{1,\const{r}}(\infty)$, and
$E_{\const{r}}$ are fitting constants.

The third group consists of the true-secondary electrons from the
material.  Since the energy of the primary beam is transferred to the
secondary electrons through a cascade, their distribution peaks at
lower energy.  However, unlike the back-scattered and rediffused
electrons, a single incoming electron can produce multiple
secondaries.  As a result, the contribution of the true secondary
electrons is more involved compared to the other populations and is
\begin{gather}\label{eq:furman_ts}
  \pfrac{\gamma_{\const{ts}}}{E}(E,E,\mu') = \sum_{n=1}^M
  \frac{nP_{n,\const{ts}}(E',\mu')(E/\epsilon_n)^{p_n-1}
    \exp(-E/\epsilon_n)}{\epsilon_n\Gamma(p_n)P(np_n,
    E'/\epsilon_n)}P\big((n-1)p_n,(E'-E)\epsilon_n\big),
  \\ P_{n,\const{ts}}(E',\mu') = \binom{M}{n} \left(\frac{
    \hat{\gamma}(\mu') D\left[ E' / \hat{E}(\mu') \right]}{M}\right)^n
  \left(1- \frac{ \hat{\gamma}(\mu') D\left[ E' / \hat{E}(\mu')
      \right]}{M}\right)^{M-n},\nonumber\\ \hat{\gamma}(\mu') =
  \hat{\gamma}_{\const{ts}} \left[1 + t_1\left(1 - \mu'^{t_2}\right)
    \right], \quad \hat{E}(\mu') = \hat{E}_{\const{ts}} \left[1 +
    t_3\left(1 - \mu'^{t_4}\right) \right],\nonumber\\ D(x) =
  \frac{sx}{s-1+x^s},\nonumber
\end{gather}
where $t_1$, $t_2$, $t_3$, $t_4$, $p_n$, $\epsilon_n$,
$\hat{\gamma}_{\const{ts}}$, $\hat{E}_{\const{ts}}$ and $s$ are
fitting variables. $\Gamma(\cdot)$ is the gamma function and
$P(\cdot,\cdot)$ is the normalized incomplete gamma
function.\footnote{$P(0,x)=1$} Note that the summation in
\eqr{furman_ts} should theoretically go to infinity, but error from
limiting it to $M=10$ is negligible \cite{Furman2002}.

An example of secondary electron distributions emitted by a single
\SI{200}{eV} electron beam calculated from \eqr{furman_e},
\eqr{furman_r}, and \eqr{furman_ts} using parameters from Tab.~I and
Tab.~II of Ref. \cite{Furman2002} are in \fgr{furman}.  Integrating
the area under the curves of individual populations yields the total
gains $\gamma_{\const{e}}=0.1241$ for back-scattered electrons,
$\gamma_{\const{r}}=0.7350$ for rediffused, and
$\gamma_{\const{ts}}=1.1283$ for true-secondary electrons for
stainless steel. Note that $\gamma_{\const{e}}+\gamma_{\const{r}} <1$
is required but, for this case, $\gamma_{\const{ts}} > 1$. Note that
even though more particles are returned to the system, their total
kinetic energy is lower than the energy of the incoming beam.
\begin{figure}[!htb]
  \centering
  \includegraphics[width=0.8\linewidth]{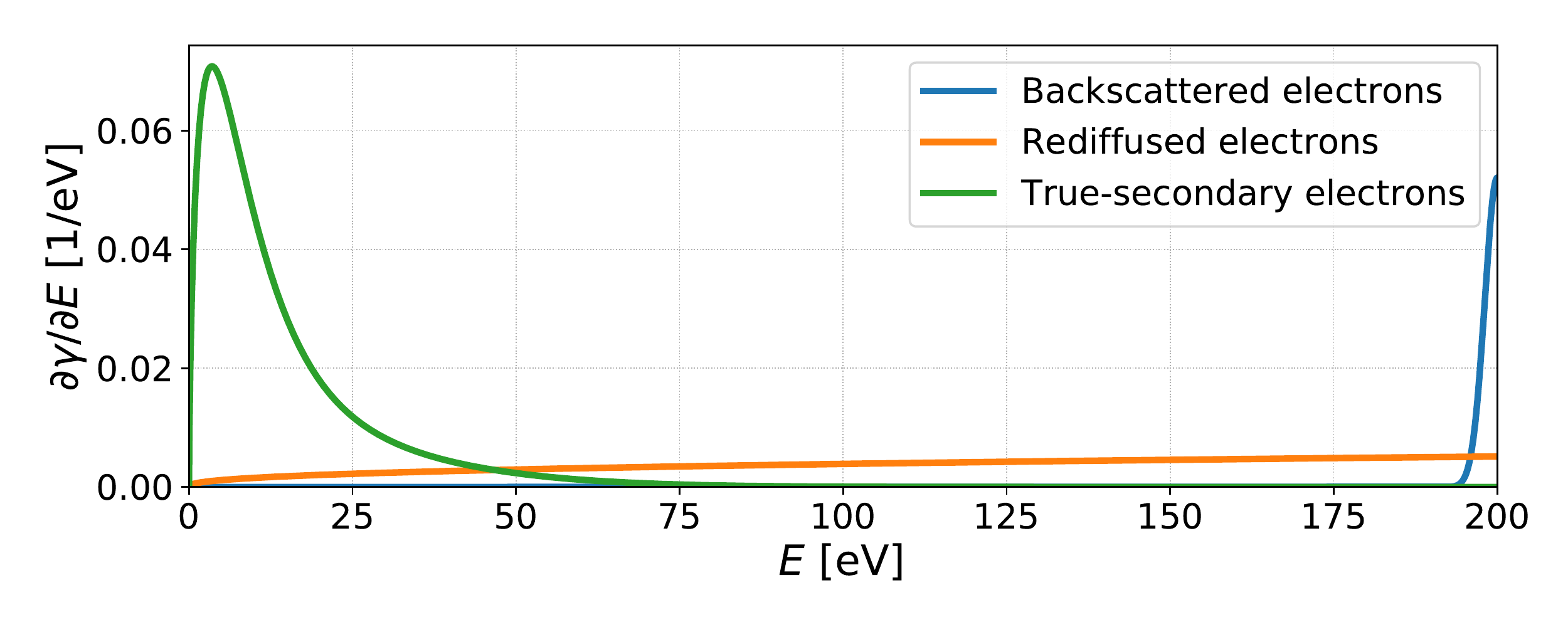}
  \caption{The energetic distribution of the three particle
    populations emitted by a single \SI{200}{eV} electron
    mono-energetic beam with normal incidence for stainless steel
    based on the phenomenological model fits \cite{Furman2002}. Figure
    taken from a Ph.D dissertation \cite{Cagas2018}.}
  \label{fig:furman}
\end{figure}

The gain, $\partial\gamma/\partial E$, is required as an intermediate
step to obtain emission probabilities for Monte-Carlo SEE codes.
However, $\partial\gamma/\partial E'(E,E',\mu)$ resembles the
reflection function in \eqr{genbc} since
\begin{equation*}
  \int_0^\infty \pfrac{\gamma}{E}\,\dif E = \gamma(E').
\end{equation*}
The dependence on the outgoing angle is the only missing part.
Experimental measurements show that the dependence is a cosine
function for the true-secondary electrons \citep{Bruining1954}, i.e.,
the incoming and outgoing angles are completely uncorrelated.  While
this is not quite true for the other two populations, Furman \& Pivi
\cite{Furman2002} make this assumption as well.  In other words, this
model can be directly used as a reflection function, $R$, in
\eqr{genbc},
\begin{equation*}
  f_{\text{out}}(E,\mu) = \int_0^1\int_0^\infty
  \mu \pfrac{\gamma}{E}(E,E',\mu')f_{\text{in}}(E',\mu') \,\dif E'\dif\mu'.
\end{equation*}
The integral can be seen as a ``summation'' over all incoming cold
beams in order to extend the mono-energetic formulation for thermal
populations.  Finally, the expression needs to be correctly
transformed from energetic units, typical for surface physics, to
phase space velocity coordinates.  Noting that
\begin{equation*}
  \pfrac{\gamma}{v_{\const{x}}} =
  \pfrac{\gamma}{E}\pfrac{E}{v_{\const{x}}} =
  \pfrac{\gamma}{E}mv_{\const{x}},
\end{equation*}
the boundary condition 1D ($\mu \equiv 1$) becomes
\begin{equation}
  f_{\text{out}}(v_{\const{x}}) = \int \mu(v_{\const{x}})
  \pfrac{\gamma}{E}\big(E(v_{\const{x}}),E(v_{\const{x}}'),\mu(v_{\const{x}}')\big)
  mv_{\const{x}} f_{\text{in}}(v_{\const{x}}')\,\dif v_{\const{x}}'.
\end{equation}

The behavior of this reflection function can be tested on a synthetic
Maxwellian distribution function.  \fgr{furman_bc} presents the
results with colors of the populations corresponding to those in
\fgr{furman}.  Note that the reflected electrons have the largest
population when using a distribution, even though their gain in
\fgr{furman} is the smallest.  \fgr{furman} depicts a case where
incoming particles have enough energy to penetrate the material,
limiting the contribution of the back-scattered electrons.  As the
energy decreases, the back-scattered electrons become the dominant
species.  This is the case for electron populations with temperatures
on the order of an electron volt with bulk velocity comparable to
thermal velocity.
\begin{figure}[!htb]
  \centering
  \includegraphics[width=0.8\linewidth]{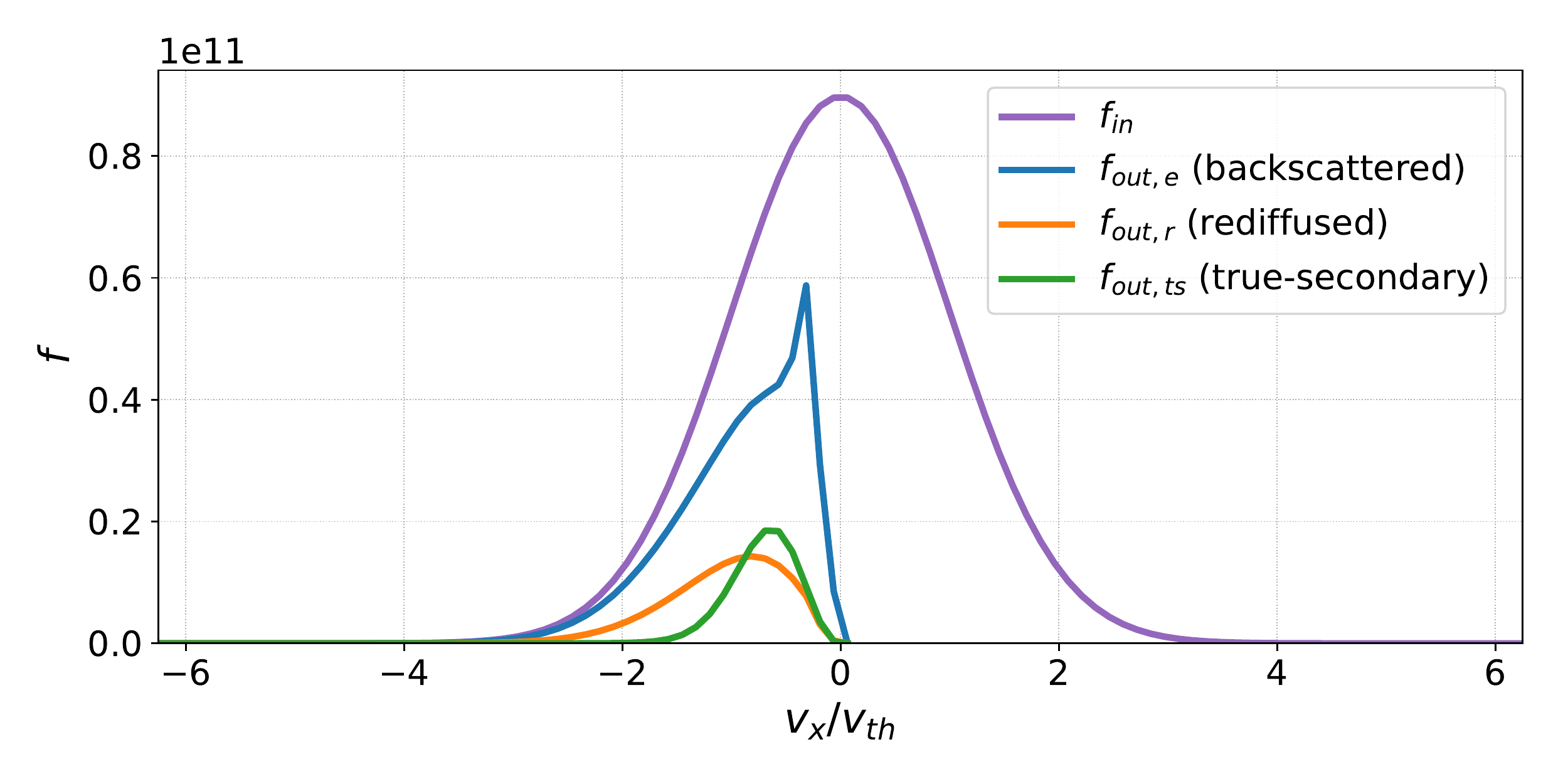}
  \caption{Application of the reflection function from
    \cite{Furman2002} on a Maxwellian distribution function.  The
    violet line represents a simulated incoming distribution function
    at the right wall boundary and blue, orange, and green are
    distributions of the reflected populations (colors correspond to
    \fgr{furman}). Figure taken from a Ph.D. dissertation
    \cite{Cagas2018}.}
  \label{fig:furman_bc}
\end{figure}

\fgr{furman_scan} provides further insight into the individual
secondary populations based on the incoming beam energy.  It extends
\fgr{furman} to include multiple incoming beam energies, i.e., the
$y$-axis of \fgr{furman_scan} corresponds to $x$-axis of \fgr{furman}.
However, since the outgoing energies are limited by the incoming
energy, the $y$-axis of \fgr{furman_scan} is normalized to the
incoming energy for better visualization.  Analogously, the values of
$\pfracb{\gamma}{E}$ are multiplied by $E'$ to allow for comparison of
magnitudes.\footnote{Since $\int_0^{E'}(\pfracb{\gamma}{E})\,\dif
  E=\gamma(E')$, normalization $(\pfracb{\gamma}{E})E'$ allows to
  compare the individual energy distributions.  Note that
  theoretically $\pfracb{\gamma}{E}\rightarrow\infty$ for
  $E'\rightarrow0$.}  This reveals a gradually decreasing relative
contribution of the true-secondary emission, while the rediffused
electron contribution remains approximately constant for a larger
range of energies prior to decreasing for $E'<\SI{20}{eV}$.  On the
other hand, as the incoming energy decreases, the backscattered
electron population becomes more significant which is consistent with
results in \fgr{furman_bc}.
\begin{figure}[!htb]
  \centering
  \includegraphics[width=0.8\linewidth]{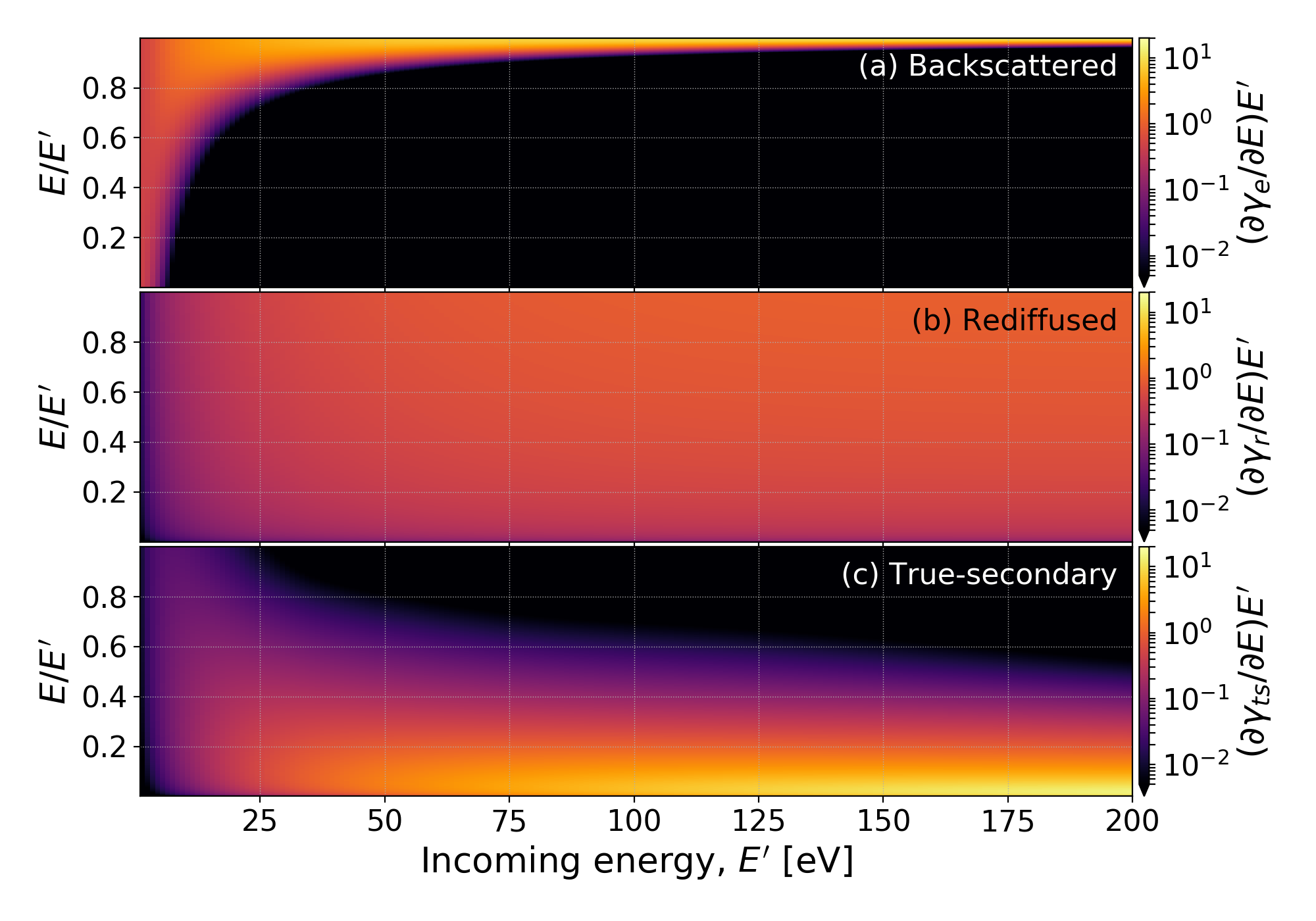}
  \caption{Contributions of the secondary populations from the
    phenomenological model \cite{Furman2002}, based on the incoming
    energy.  Both the values and the $y$-axis are normalized to the
    incoming energy to allow for better comparison of the relative
    contributions.  From top to bottom the figure captures
    backscattered (elastically reflected) electrons, rediffused
    electrons, and true-secondary electrons. Figure taken from a Ph.D.
    dissertation \cite{Cagas2018}.}
  \label{fig:furman_scan}
\end{figure}

Furthermore, note that although the model is mathematically sound for
incoming energies all the way to zero, the values at the lower energy
range, which are crucial as described above, are from an extrapolation
of higher energy beam data.  Therefore, for simulating $\sim
\SI{10}{eV}$ electron distributions in contact with a wall, a
different model specifically tailored for these energies might be
preferred.  Another disadvantage of the model is its dependency on a
significant number of fitting parameters which do not necessarily
correspond to physical quantities.  This limits the materials that
this model extends to.

\subsection{Electron Emission Boundary Condition: Bronold \& Fehske (2015) model}

Bronold \& Fehske \cite{Bronold2015} present a model for electron
absorption by a dielectric wall, which is based on first principles
from quantum mechanics.  In comparison to the previous model, it has
fewer fitting parameters and most of them have physical relevance.  As
a result, it is applicable to a wide range of materials based on
standard physical parameters that can be obtained from material
databases.  The disadvantage of this model is that it is accurate only
up to incoming energies comparable to the electron band gap
$E_{\const{g}}\sim\SI{10}{eV}$ ($E_{\const{g}}=\SI{7.8}{eV}$ for MgO
used for examples here).

Here, the reflection function is defined directly,
\begin{equation}\label{eq:bronold}
  R(E,\mu,E',\mu') =
  \underbracket{R(E',\mu')\delta(E-E')\delta(\mu-\mu')}_{\text{backscattered}}
  + \underbracket{\delta R(E,\mu,E',\mu')}_{\text{rediffused}}.
\end{equation}
Note that the model assumes specular reflection for the back-scattered
electrons, i.e., the energy and angles are conserved with variable
probability $R(E',\mu')$, which is a function only of the incoming
properties.  It is given as $R(E',\mu') = 1 - \mathcal{T}(E',\mu')$,
where $\mathcal{T}(E',\mu')$ is the probability of a
quantum-mechanical reflection,
\begin{equation*}
  \mathcal{T}(E',\mu') =
  \frac{4\overline{m}_{\const{e}}kp}{(\overline{m}_{\const{e}}k+p)^2}, \quad k =
  \sqrt{E'-\chi}\mu', \quad p=\sqrt{\overline{m}_{\const{e}}E'}\nu'.
\end{equation*}
$\overline{m}_{\const{e}}$ is the relative mass of a conduction band
electron and $\chi$ is the electron affinity of the dielectric.  $k$
and $p$ are components of momentum perpendicular to the wall where
$\nu$ is the cosine angle inside the wall.  $\nu$ is connected with
$\mu$ through conservation of energy and lateral momentum,
\begin{equation}\label{eq:bronold_conservation}
  1-\nu'^2 = \frac{E'-\chi}{\overline{m}_{\const{e}}E'}(1-\mu'^2).
\end{equation}

The probability of reflection, $R(E',\mu')$ is captured in
\fgr{bc_R_2D}.  Note the region of $R(E',\mu')= 1$ for
$E'<\SI{1}{eV}$.  This region contains electrons with energy below the
affinity of the material.  These electrons cannot penetrate the
potential barrier and are all reflected.  As a direct consequence of
the conservation of energy and lateral momentum
\eqrp{bronold_conservation}, there is a critical angle given as
$\mu_{\const{c}} = \sqrt{1-\overline{m}_{\const{e}}E'/(E'-\chi)}$.
Particles entering under this angle have the momentum vector
perpendicular to the surface after penetrating the material; particles
that hit the wall with $\mu'<\mu_{\const{c}}$ are reflected.  Note
that particles with $\mu'>\mu_{\const{c}}$ and $E'>\SI{2}{eV}$
generally do penetrate the material and would be lost from the plasma
if back-scattering was the only effect taken into account. They can,
however, return to the plasma through rediffusion.

\begin{figure}[!htb]
  \centering
  \includegraphics[width=0.8\linewidth]{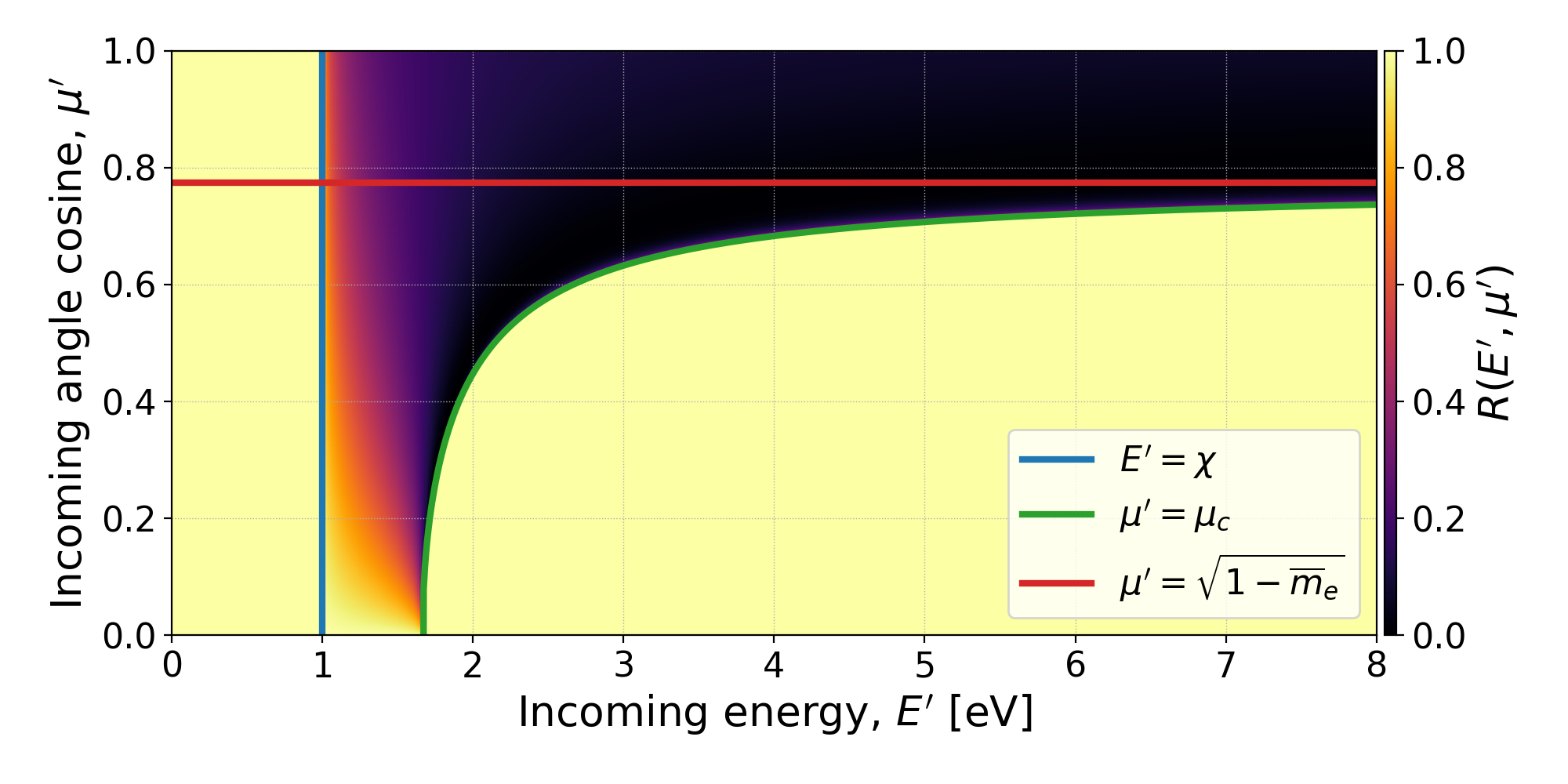}
  \caption{Probability of back-scattering, $R(E',\mu')$, from the
    quantum mechanics based model \cite{Bronold2015} as a function of
    incoming angle and energy. Highlighted are $E'=\chi$ (blue line)
    below which all particles are reflected, and the critical angle
    $\mu_{\const{c}}$ (green line) given by the conservation laws,
    \eqrp{bronold_conservation}.  The red line marks the angle above
    which rediffusion is possible, \eqrp{bronold_rediffusion}.  Used
    parameters are for MgO, $\chi=\SI{1}{eV}$ and
    $\overline{m}_{\const{e}}=0.4$. Figure adapted from a
    Ph.D. dissertation \cite{Cagas2018}.}
  \label{fig:bc_R_2D}
\end{figure}

Description of rediffusion is more involved in comparison to
back-scattered electrons \cite{Bronold2018},
\begin{equation}\label{eq:bronold_rediffusion}
  \delta R(E,\mu,E',\mu')=\pfrac{\nu}{\mu} \mathcal{T}(E',\mu')
  \rho(E) \mathcal{B}(E,\mu,E',\mu') \mathcal{T}(E,\mu)
  \theta\big(\mu-\sqrt{1-\overline{m}_{\const{e}}}\big),
\end{equation}
where $\rho(E) = \sqrt{\overline{m}_{\const{e}}^3E}/2(2\pi)^3$ is the
conduction band density of states and
\begin{equation*}
  \mathcal{B}(E,\mu,E',\mu') =
  \frac{Q(E,\mu,E',\mu')}{\int_0^1\int_0^{E'}\rho(E)Q(E,\mu,E',\mu')\,\dif
    E\dif\mu}
\end{equation*}
is the probability of rediffusion.  $Q(E,\mu,E',\mu')$ is given by a
recursive relation summed over the back-scattering events inside the
material.  Note the Heaviside step function in
\eqr{bronold_rediffusion}; the limiting $\mu$ is marked by the blue
line in \fgr{bc_R_2D}.  The population with cosine angles above this
line can return to the domain after penetrating the material,
significantly influencing \eqr{bronold}.  True-secondary electrons
excited by incoming electrons with energies considered here
($<\SI{10}{eV}$) are neglected in this model.\footnote{Bronold \&
  Fehske (2018; \cite{Bronold2018}) also discuss true-secondary
  electrons excited with energy coming change of internal energy
  levels of incoming ions; these effects are neglected in this work.}

This model can be implemented into the simulation in the same manner
as the previous phenomenological model \cite{Furman2002}.  All the
relations above are derived for ideally flat walls without any
defects.  To address effects of real walls, the authors modify the
relations by adding a roughness parameter $C$ \cite{Smith1998}, which
is proportional to the density scattering centers.  With $C=1$ and
$C=2$ the results match experimental data very well (see
Fig.\thinspace3 in \cite{Bronold2015}; results are much better than
for $C=0$).  Furthermore, with increasing $C$, the effects of $\delta
R(E,\mu,E',\mu')$ become less important.  This presents an opportunity
to develop reasonably accurate and computational inexpensive boundary
conditions by neglecting the rediffusion and using the
roughness-modified formula for the probability of a quantum-mechanical
reflection (Eq.\thinspace(13) in \cite{Bronold2015}),
\begin{equation}\label{eq:bronold_C}
  \overline{\mathcal{T}}(E',\mu') =
  \frac{\mathcal{T}(E',\mu')}{1+C/\mu'} -
  \frac{C/\mu'}{1+C/\mu'}\int_{\mu_{\const{c}}}^1\mathcal{T}(E',\mu'')\,\dif\mu''.
\end{equation}
Calculating these integrals for the reflection function of each
particle would still be quite expensive.  However, as emphasized
before, the energies and angles need to be treated as coordinates and
the integrals can be precomputed substantially decreasing
computational cost.

The reflection function, $R$, calculated with $\overline{\mathcal{T}}$
then significantly alters \fgr{bc_R_2D}.  The modified version is in
\fgr{bc_Rt_2D}.  Particularly noticeable is the absence of regions
with absolute reflection in the bottom-right sector (higher energies
and oblique angles).
\begin{figure}[!htb]
  \centering
  \includegraphics[width=0.8\linewidth]{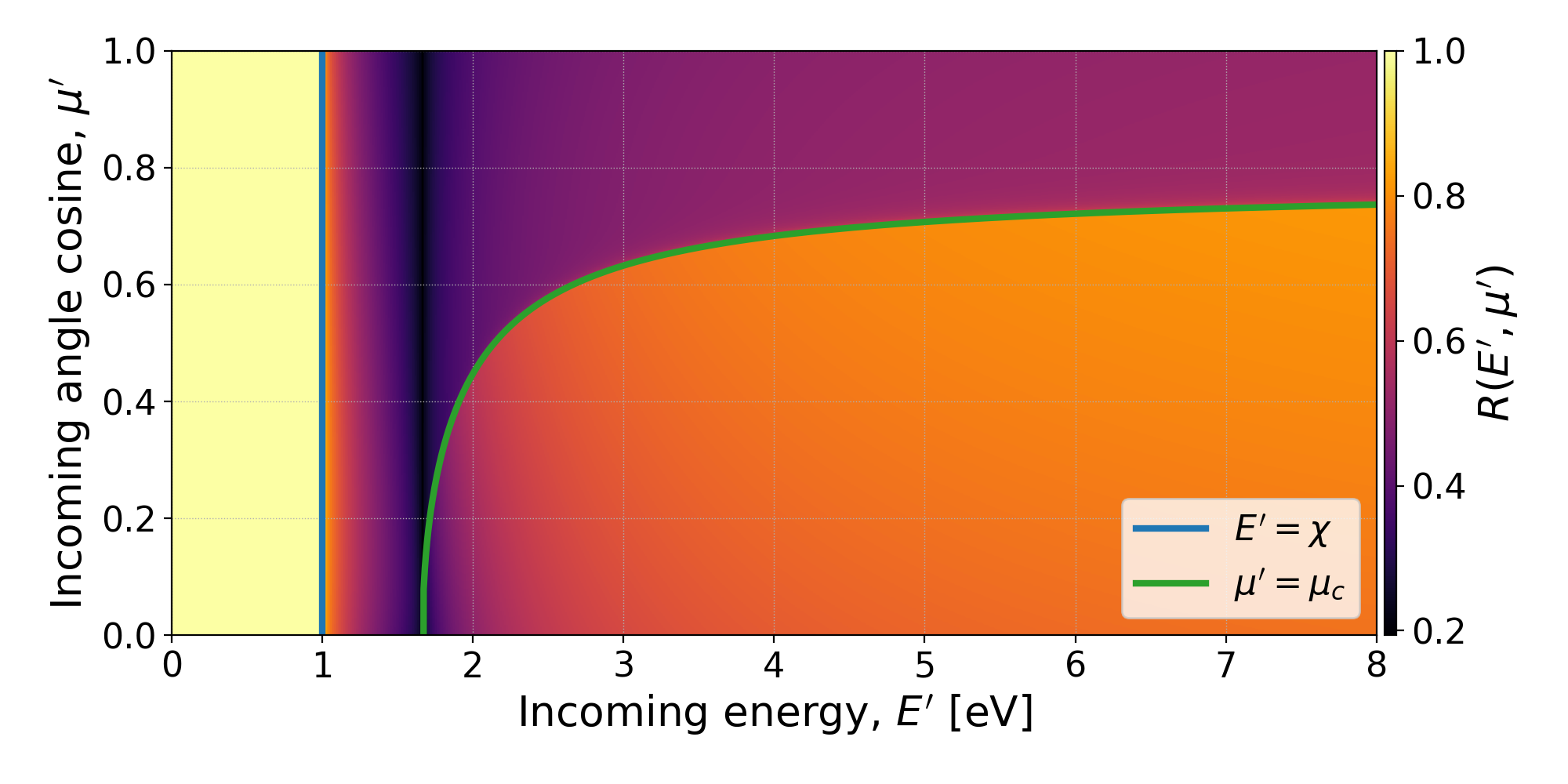}
  \caption{Probability of back-scattering, $R(E',\mu')$
    \cite{Bronold2015} modified with the roughness coefficient $C$
    \eqrp{bronold_C}.  Using $C=2$ and material parameters for MgO
    ($\chi=\SI{1}{eV}$ and $\overline{m}_{\const{e}}=0.4$). Figure
    taken from a Ph.D.  dissertation \cite{Cagas2018}.}
  \label{fig:bc_Rt_2D}
\end{figure}

The whole process can be performed as follows.  The
reflection function is defined as
\begin{equation}\label{eq:bronold_R}
  R(E,\mu,E',\mu') = \left(1 - \frac{\mathcal{T}(E',\mu')}{1+C/\mu'} -
  \frac{C/\mu'}{1+C/\mu'}\int_{\mu_c}^1\mathcal{T}(E',\mu'')\,\dif\mu''\right)
  \delta(E-E')\delta(\mu-\mu').
\end{equation}
\eqr{bronold_R} is converted from energetic to velocity coordinates
and then substituted into the general formula in \eqr{genbc} and the
integration over $\bm{v}'$ is performed, which is made simple by the
Dirac delta functions.

The following section (\ref{sec:dg}) provides an example of
implementation into the discontinuous Galerkin continuum kinetic model
of the \texttt{Gkeyll} simulation framework.

\section{Applications for Discontinuous Galerkin Simulations}\label{sec:dg}

The continuum kinetic boundary condition descriptions presented in
this work are independent of the choice of the numerical method.
Here, the discontinuous Galerkin (DG) scheme is used to develop and
apply the boundary conditions described.  The DG method is
advantageous as it allows an arbitrarily high order representation of
the solution and a small stencil size regardless of spatial order
\cite{Reed1973,Cockburn2001,Hesthaven2004}.  This section describes
the implementation in the \texttt{Gkeyll} framework (see the
\ref{app:gke} for instruction on how to get \texttt{Gkeyll}) and the
first self-consistent continuum kinetic simulation results. DG
discretization in \texttt{Gkeyll} is based on a novel matrix-free
algorithm [publication in preparation], which makes it well suited for
the implementation of phenomenological and first-principles models.
The Vlasov-Maxwell solver has been used in previous work for various
problems like plasma-material interactions \cite{Cagas2017s,Cagas2018}
and astrophysically relevant problems
\cite{Cagas2017c,Juno2018,Pusztai2018}.

The DG scheme is developed using the principle of weak equality.
\begin{definition}
  Two functions, $f$ and $g$, are weakly equal if
  \begin{equation}
    \langle \psi,f-g \rangle = 0,
  \end{equation}
  where $\langle\cdot,\cdot\rangle$ is an inner product and $\psi$ is
  a test function. The weak equality will be denoted as $f\circeq g$.
\end{definition}
In \texttt{Gkeyll}, the inner product is defined as
\begin{equation}
  \langle \psi, f\rangle = \int_I f\psi \thinspace dz,
\end{equation}
where $z$ corresponds to a general phase space coordinate.

The evolution equations, the Vlasov and Maxwell's equations, need to
be satisfied weakly for the DG scheme.  An integration by parts is
performed so the real solution can be replaced by a polynomial
representation, $f_{\const{h}} = \sum_k \hat{f}_k\psi_k$, where
$\hat{f}_k$ are expansion coefficients corresponding to the basis
function $\psi_k$.  The Vlasov equation is then written as
\begin{align}\label{eq:weak}
  \pfrac{\widehat{f}_{n}^j(t)}{t} =
  \left(\mathcal{M}^j_{nt}\right)^{-1}\Big[\underbracket{\widehat{f}_m^j(t)
      \int_{K^j} \bm{\alpha}_{\const{h}}^j(\bm{z}) \cdot
      \big(\psi_m(\bm{z}) \nabla_{\bm{z}} \psi_t(\bm{z})\big)
      \,\dif\bm{z}}_{\text{volume term}}
    -\underbracket{\oint_{\partial K^j} \bm{F}^j(t,\bm{z})
      \psi_t(\bm{z}) \cdot \dif\bm{A}}_{\text{surface term}}\Big],
\end{align}
where $\mathcal{M}$ is a mass matrix, $\mathcal{M}^j_{nt}=\int_{K^j}
\psi_n\psi_t\,d\bm{z}$, and $\bm{\alpha}$ is a phase space vector,
$\bm{\alpha}=(\bm{v},q/m(\bm{E}+\bm{v}\times\bm{B})$. Note that the
integral in the volume term is only performed locally over cell
$K^j$. The surface term is integrated over the edge of cell $K^j$ and
includes the numerical flux function $\bm{F}$, which is a function of
$f$ in the cell $K^j$ and $f$ in its neighbours.  While the solution
is allowed to be discontinuous at each cell edge, the surface term
ties together an otherwise discontinuous representation.

At the domain boundaries, the surface terms could be used by
prescribing the numerical flux directly at the boundaries or by
including an additional layer of cells to calculate the flux in the
same manner as inside the domain.  These additional layers of cells
are typically referred to as ghost cells.  \texttt{Gkeyll} uses ghost
cells at domain boundaries.  To apply the described boundary
conditions in the ghost cells, \eqr{genbc} becomes
\begin{equation}
  f_{\const{h}}^g(\bm{x}_{\text{wall}},\,\bm{v}) = \sum_{s}
  \int_{\mathcal{V}_{\text{in}}^s} R^{gs}(\bm{v},\,\bm{v}')
  f_{\const{h}}^{s}(\bm{x}_{\text{wall}},\,\bm{v}') \,\dif\bm{v}',
\end{equation}
where the outgoing distribution function is set in the ghost layer,
$f_{\const{h}}^g$, through integration of the distribution function in
the last layer of cells in the domain next to the wall, usually
referred to as a skin layer, $f_{\const{h}}^s$. Note that in the
discrete case the integration is limited to one skin cell,
$\mathcal{V}_{\text{in}}^s$, and the contributions of these integrals
are added over the whole skin layer.  The time dependence of the
distribution functions is dropped for clarity.

Since \eqr{genbc} is defined only for
$\bm{x}=\bm{x}'=\bm{x}_{\text{wall}}$, the discrete representation is
defined only in terms of surface basis functions $\varsigma$.
Assuming, without loss of generality, that the boundary lies in the
$x$-direction, incoming and outgoing distribution functions can be
expressed as
\begin{equation*}
  f_{\const{h}}^g(\bm{x},\,\bm{v})|_{x=x_{\text{wall}}} = \sum_k
  \widehat{f}_k^g \varsigma_k(y,z,\bm{v}), \quad
  f_{\const{h}}^{s}(\bm{x},\,\bm{v}')|_{x=x_{\text{wall}}} = \sum_l
  \widehat{f}_l^{s} \varsigma_l(y,z,\bm{v}').
\end{equation*}
In the discrete weak sense, \eqr{genbc} becomes
\begin{equation*}
  \sum_k \widehat{f}_k^g \varsigma_k(y,z,\bm{v}) \circeq \sum_{s}
  \sum_l \widehat{f}_l^s \int_{\mathcal{V}_{\text{in}}^{s}}
  R_{\const{x}}^{gs}(\bm{v},\,\bm{v}') \varsigma_l(y,z,\bm{v}')
  \,\dif\bm{v}'.
\end{equation*}
The full equality is
\begin{multline}\label{eq:weakbc}
  \sum_k \widehat{f}_k^g \int_{\partial_{x} K^{g}}
  \varsigma_k(y,z,\bm{v}) \varsigma_t(y,z,\bm{v}) \,\dif y \dif z
  \dif\bm{v} =\\= \sum_{s} \sum_l \widehat{f}_l^{s} \int_{\partial_{x}
    K^{s}} \int_{\mathcal{V}_{\text{in}}^{s}}
  R_{\const{x}}^{gs}(\bm{v},\,\bm{v}')\varsigma_l(y,z,\bm{v}')
  \varsigma_t(y,z,\bm{v}) \,\dif\bm{v}' \dif y \dif z \dif\bm{v},
\end{multline}
where the phase space integrals are performed over faces of the cells
in the $x$-directions, $\partial_{x} K^{g,s}$. \eqr{weakbc} is defined
in the physical space and needs to be transformed into logical space
for numerical implementation,
\begin{multline*}
  \sum_k \widehat{f}_k^g \int_{\partial_{x} I_p}
  \widehat{\varsigma}_k(\eta_{\const{y}},\eta_{\const{z}},\bm{\eta}_{\bm{v}})
  \widehat{\varsigma}_t(\eta_{\const{y}},\eta_{\const{z}},\bm{\eta}_{\bm{v}})
  \,\dif\eta_{\const{y}} \dif\eta_{\const{z}} \dif\bm{\eta}_{\bm{v}}
  =\\= \frac{\prod_{i=1}^{d_v}\Delta v_i}{2^{d_v}} \sum_{s,l}
  \widehat{f}_l^{s} \int_{\partial_{x} I_p} \int_{I_v}
  R_{\const{x}}^{gs}\big(\bm{v}^g(\bm{\eta}_{\bm{v}}),\,\bm{v}^s(\bm{\eta}_{\bm{v}}')\big)
  \widehat{\varsigma}_l(\eta_{\const{y}},\eta_{\const{z}},\bm{\eta}_{\bm{v}}')
  \widehat{\varsigma}_t(\eta_{\const{y}},\eta_{\const{z}},\bm{\eta}_{\bm{v}})
  \,\dif\bm{\eta}_{\bm{v}}'\dif\eta_{\const{y}} \dif\eta_{\const{z}}
  \dif\bm{\eta}_{\bm{v}},
\end{multline*}
where $\Delta v$ denotes the size of the cell in each dimension.  For
simplicity, the mesh is assumed to be uniform.  For an arbitrary mesh,
the Jacobian containing geometric information for each cell would be
inside the summation operators.  An orthonormal surface basis can be
constructed, $\int_{\partial_{x} I_p}
\tilde{\varsigma}_k(\eta_{\const{y}},\eta_{\const{z}},\bm{\eta}_{\bm{v}})
\tilde{\varsigma}_t(\eta_{\const{y}},\eta_{\const{z}},\bm{\eta}_{\bm{v}})
\,\dif\eta_{\const{y}} \dif\eta_{\const{z}} \dif\bm{\eta}_{\bm{v}} =
\delta_{kt}$.  The relation simplifies to
\begin{equation}\label{eq:weakbc2}
  \widehat{f}_k^g = \frac{\prod_{i=1}^{d_v}\Delta v_i}{2^{d_v}}
  \sum_{s,l} \widehat{f}_l^{s} \int_{\partial_{x} I_p} \int_{I_v}
  R_{\const{x}}^{gs}\big(\bm{v}^g(\bm{\eta}_{\bm{v}}),\,\bm{v}^s(\bm{\eta}_{\bm{v}}')\big)
  \widehat{\varsigma}_l(\eta_{\const{y}},\eta_{\const{z}},\bm{\eta}_{\bm{v}}')
  \widehat{\varsigma}_k(\eta_{\const{y}},\eta_{\const{z}},\bm{\eta}_{\bm{v}})
  \,\dif\bm{\eta}_{\bm{v}}'\dif\eta_{\const{y}} \dif\eta_{\const{z}}
  \dif\bm{\eta}_{\bm{v}}.
\end{equation}
However, since $R^{gs}\big(\bm{v}^g(\bm{\eta}_{\bm{v}}),
\,\bm{v}^s(\bm{\eta}_{\bm{v}}')\big)$ can have a complicated
dependence on $\bm{v}$ and $\bm{v}'$, the integral on the
right-hand-side of \eqr{weakbc2} cannot usually be precomputed in the
logical space as can be done for volume and surface terms of the
Vlasov equation.  The integrals, however, do not change in time and,
therefore, can be precomputed for each wall cell during the setup
phase of a simulation.\footnote{It should be mentioned that
  \eqr{weakbc} does not exactly represent the boundary condition
  implementation in the \texttt{Gkeyll} framework.  There, the fact is
  used that the distribution function in the ghost layer is used only
  to calculate the numerical flux.  Therefore, volume basis functions,
  $\tilde{\psi}$, can be used instead of the surface ones,
  $\widehat{\varsigma}$, as long as the logical space coordinates in
  the direction perpendicular to the boundary have opposite
  signs. With the wall perpendicular to the $x$-direction,
  $\varsigma_l(\eta_{\const{y}},\eta_{\const{z}},\bm{\eta}_{\bm{v}}')$
  and
  $\varsigma_k(\eta_{\const{y}},\eta_{\const{z}},\bm{\eta}_{\bm{v}}')$
  are replaced with
  $\widehat{\psi}_l(-\eta_{\const{x}},\eta_{\const{y}},\eta_{\const{z}},\bm{\eta}_{\bm{v}}')$
  and
  $\widehat{\psi}_k(\eta_{\const{x}},\eta_{\const{y}},\eta_{\const{z}},\bm{\eta}_{\bm{v}}')$. This
  guaranties that the basis function
  $\widehat{\psi}_k|_{\eta_{\const{x}}=\pm1}$ in a ghost layer has the
  same value as the basis function
  $\widehat{\psi}_l|_{\eta_{\const{x}}=\mp1}$ in a skin layer as it
  would have been a case for the surface basis functions. Then
  integral then needs to be evaluated over the whole cell,
  $I_p$. \eqr{weakbc2} transforms into
\begin{multline}\label{eq:bc3}
  \widehat{f}_k^g = \frac{\prod_{i=1}^{d_v}\Delta v_i}{2^{d_v}}
  \sum_{s,l} \widehat{f}_l^{s} \int_{I_p} \int_{I_v}
  R_{\const{x}}^{gs}\big(\bm{v}^g(\bm{\eta}_{\bm{v}}),\,\bm{v}^s(\bm{\eta}_{\bm{v}}')\big)
  \\\widehat{\psi}_l(-\eta_{\const{x}},\eta_{\const{y}},\eta_{\const{z}},\bm{\eta}_{\bm{v}}')
  \widehat{\psi}_k(\eta_{\const{x}},\eta_{\const{y}},\eta_{\const{z}},\bm{\eta}_{\bm{v}})
  \,\dif\bm{\eta}_{\bm{v}}'\dif\eta_{\const{x}} \dif\eta_{\const{y}}
  \dif\eta_{\const{z}} \dif\bm{\eta}_{\bm{v}}.
\end{multline}}

\subsection{Specular Reflection}
As a proof of concept, the specular reflection from \eqr{specular} is
implemented. In the discrete case, the reflection function is limited
only to cells with ``opposite $x$-velocity'', symbolically denoted
with Kronecker delta,\footnote{Note there is a difference between the
  Kronecker delta, $\delta_{ij}$, and the Dirac delta function,
  $\delta(x)$.}
\begin{equation}
  R_{\const{x}}^{gs}(\bm{v},\bm{v}') = \delta_{g(-s)}
  \delta(v_{\const{x}}+v_{\const{x}}')\delta(v_{\const{y}}-v_{\const{y}}')\delta(v_{\const{z}}-v_{\const{z}}').
\end{equation}
Substituting this into \eqr{bc3} yields
\begin{equation}\label{eq:bc4}
  \widehat{f}_k^g = \sum_{l} \widehat{f}_l^{-s} \int_{I_p}
  \widehat{\psi}_l(-\eta_{\const{x}},\eta_{\const{y}},\eta_{\const{z}},-\eta_{v_{\const{x}}},\eta_{v_{\const{y}}},
  \eta_{v_{\const{z}}})
  \widehat{\psi}_k(\eta_{\const{x}},\eta_{\const{y}},\eta_{\const{z}},\eta_{v_{\const{x}}},\eta_{v_{\const{y}}},
  \eta_{v_{\const{z}}}) \,\dif\bm{\eta}_{\bm{x}}
  \dif\bm{\eta}_{\bm{v}}.
\end{equation}

For a 1X1V (one dimension in configuration space and one dimension in
velocity space) simulation with polynomial order 2, \texttt{Gkeyll}
uses the following Serendipity \cite{Arnold2011} basis functions
\begin{align}\begin{aligned}\label{eq:1x1v}
  \widehat{\psi}_0(\eta_{\const{x}},\eta_{v_{\const{x}}}) &=
  \frac{1}{2} &
  \widehat{\psi}_1(\eta_{\const{x}},\eta_{v_{\const{x}}}) &=
  \frac{\sqrt{3}\eta_{\const{x}}}{2}
  \\ \widehat{\psi}_2(\eta_{\const{x}},\eta_{v_{\const{x}}}) &=
  \frac{\sqrt{3}\eta_{v_{\const{x}}}}{2} &
  \widehat{\psi}_3(\eta_{\const{x}},\eta_{v_{\const{x}}}) &=
  \frac{3\eta_{\const{x}}\eta_{v_{\const{x}}}}{2}
  \\ \widehat{\psi}_4(\eta_{\const{x}},\eta_{v_{\const{x}}}) &=
  \frac{\sqrt{5}(3\eta_{\const{x}}^2-1)}{4} &
  \widehat{\psi}_5(\eta_{\const{x}},\eta_{v_{\const{x}}}) &=
  \frac{\sqrt{5}(3\eta_{v_{\const{x}}}^2-1)}{4}
  \\ \widehat{\psi}_6(\eta_{\const{x}},\eta_{v_{\const{x}}}) &=
  \frac{\sqrt{15}(3\eta_{\const{x}}^2-1)\eta_{v_{\const{x}}}}{4} &
  \widehat{\psi}_7(\eta_{\const{x}},\eta_{v_{\const{x}}}) &=
  \frac{\sqrt{15}\eta_{\const{x}}(3\eta_{v_{\const{x}}}^2-1)}{4}.
\end{aligned}\end{align}
Due to the orthonormality of the basis functions, the boundary
condition is reduced to a simple matrix operation,
\begin{gather*}
  \mathcal{R}_{kl} = \int_{I_p}
  \widehat{\psi}_l(-\eta_{\const{x}},-\eta_{v_{\const{x}}})
  \widehat{\psi}_k(\eta_{\const{x}},\eta_{v_{\const{x}}})
  \,\dif\eta_{x} \dif\eta_{v_{\const{x}}},\\ = \begin{pmatrix} 1 & 0 &
    0 & 0 & 0 & 0 & 0 & 0 \\ 0 & -1 & 0 & 0 & 0 & 0 & 0 & 0 \\ 0 & 0 &
    -1 & 0 & 0 & 0 & 0 & 0 \\ 0 & 0 & 0 & 1 & 0 & 0 & 0 & 0 \\ 0 & 0 &
    0 & 0 & 1 & 0 & 0 & 0 \\ 0 & 0 & 0 & 0 & 0 & 1 & 0 & 0 \\ 0 & 0 &
    0 & 0 & 0 & 0 & -1 & 0 \\ 0 & 0 & 0 & 0 & 0 & 0 & 0 & -1
  \end{pmatrix}.
\end{gather*}

\fgr{bounce} presents a simple example when this boundary condition is
used to let an originally Gaussian distribution of neutral particles
bounce between two walls.
\begin{figure}[!htb]
  \centering
  \includegraphics[width=0.8\linewidth]{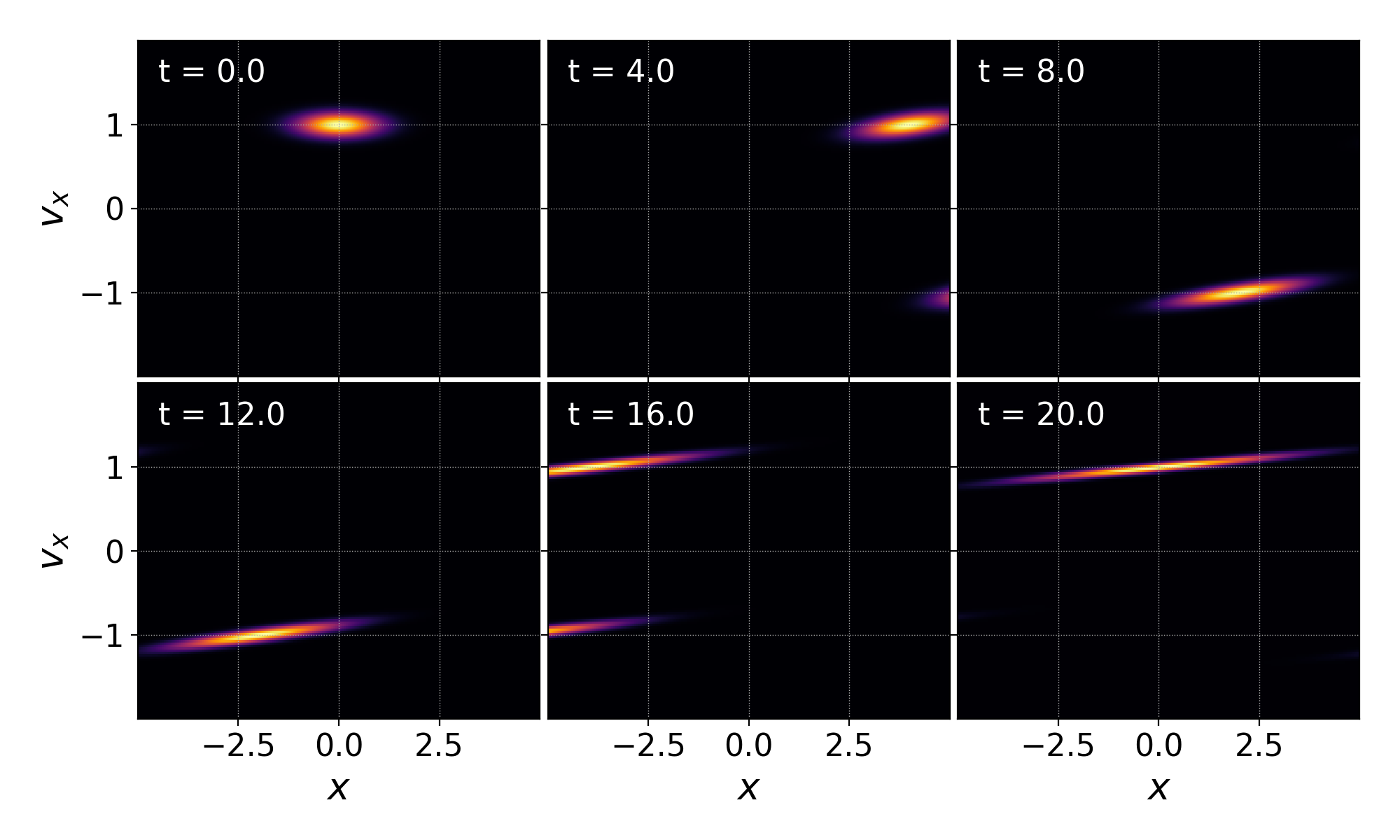}
  \caption{A test problem for the general boundary condition framework
    with implementation of specular reflection. An initially Gaussian
    distribution of neutral particles bounce between two walls in a 1D
    domain.}
  \label{fig:bounce}
\end{figure}
This boundary condition can be used to save computation time for
symmetric problems.  Plasma sheath simulations without magnetic fields
where walls bound the plasma on both sides of the domain are an
example of symmetric problems. Instead of simulating walls on both
sides, the domain could be cut in half using a wall boundary on one
side and a specular reflection to represent symmetry for the other
boundary. To test this, a full domain simulation using two absorbing
wall boundaries is compared with a half domain simulation which uses
the specular boundary condition to capture the symmetries.  The
results are presented in \fgr{bc_reflect}.  Note that the values of
the distribution functions are directly subtracted and that the figure
shows only the right half of the full domain simulation to allow
direct calculation of the difference.  Since there are regions where
the distribution function is close to zero, the difference is
normalized to the maximum value of the distribution.  The relative
difference on the order of $10^{-13}$ represents accumulated round-off
error which verifies that this boundary condition implementation is
correct.
\begin{figure}[!htb]
  \centering
  \includegraphics[width=0.8\linewidth]{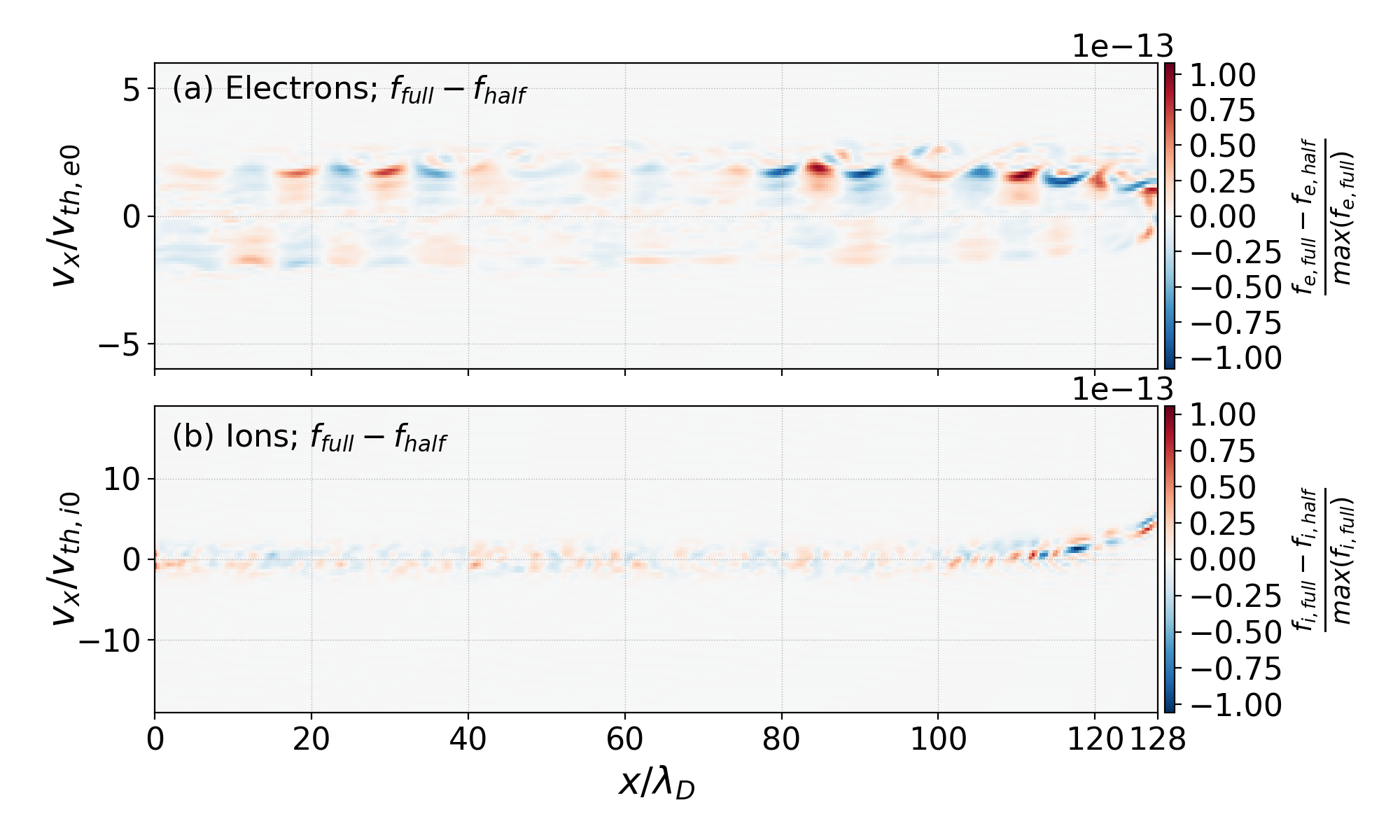}
  \caption{Normalized difference between distribution functions from
    the full domain (only right half is plotted) simulation using
    ideally absorbing walls on both sides and the half domain
    simulation with the specular boundary condition at the left edge
    replicating the symmetric behavior. Figure taken from a Ph.D.
    dissertation \cite{Cagas2018}.}
  \label{fig:bc_reflect}
\end{figure}

\subsection{Dielectric Boundary Condition for Electrons}

The boundary condition for electrons based on the model by Bronold \&
Fehske \cite{Bronold2015} is implemented. The integral which needs to
be solved is given as
\begin{multline}
  \mathcal{R}_{\const{x},kl}^g = \int_{I_p} \left(1 -
  \frac{\mathcal{T}\big(E^g(\bm{\eta}_{\bm{v}}),
    \mu^g(\bm{\eta}_{\bm{v}})\big)}{1+C/\mu^g(\bm{\eta}_{\bm{v}})} -
  \frac{C/\mu^g(\bm{\eta}_{\bm{v}})}{1+C/\mu^g(\bm{\eta}_{\bm{v}})}
  \int_{\mu_{\const{c}}^g(\bm{\eta}_{\bm{v}})}^1\mathcal{T}(E^g(\bm{\eta}_{\bm{v}}),
  \mu'')\,\dif\mu''\right)\times
  \\\widehat{\psi}_l(-\eta_{\const{x}},\eta_{\const{y}},\eta_{\const{z}},-\eta_{v_{\const{x}}},\eta_{v_{\const{y}}},\eta_{v_{\const{z}}})
  \widehat{\psi}_k(\eta_{\const{x}},\eta_{\const{y}},\eta_{\const{z}},\eta_{v_{\const{x}}},\eta_{v_{\const{y}}},\eta_{v_{\const{z}}})
  \,\dif\eta_{\const{x}} \dif\eta_{\const{y}}
  \dif\eta_{\const{z}}\dif\eta_{v_{\const{x}}}
  \dif\eta_{v_{\const{y}}} \dif\eta_{v_{\const{z}}}.
\end{multline}
Due to the complexity of $R$ the boundary condition needs to be
precomputed for each cell.  As seen in \fgr{bc_R_2D}, $R = 1$ for low
energies and decreases fairly rapidly as the energy increases.
Therefore, it is important to be careful with constructing the
velocity mesh.  The electron mesh used for previous simulations
\cite{Cagas2017s} extends from $-6\,v_{th,e}$ to $6\,v_{th,e}$ and
uses 32 cells. This puts the sharp transition at $E'=\chi$ inside the
second cell (counting from the center).  As the polynomial
approximation is not suited for such sharp transitions, projection of
$R$ onto this mesh results in significant overshoot; see blue line in
\fgr{bc_R}.  However, noting the ability of the DG method to handle
discontinuities and sharp gradients between the cells, the velocity
mesh can be tailored for the purposes of the boundary condition.  As
seen by the orange line in \fgr{bc_R_2D}, tailoring the mesh
eliminates the overshoot at $v_{\const{x}}\approx 0.5\,v_{th}$.
\begin{figure}[!htb]
  \centering
  \includegraphics[width=0.8\linewidth]{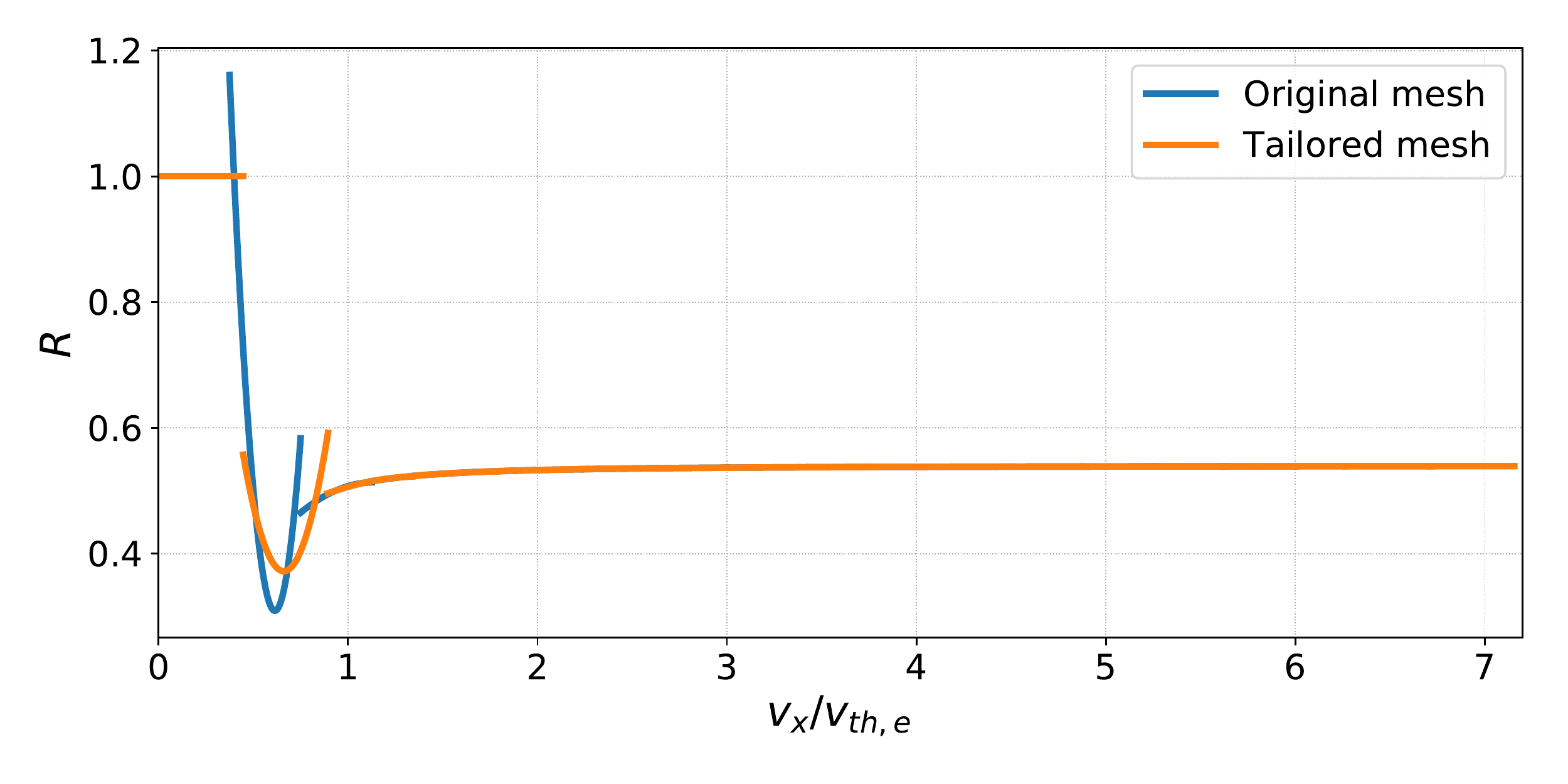}
  \caption{An example showing the projection of the reflection
    function given by \eqr{bronold_R} onto the simulation mesh.  The
    same mesh is used as with previous simulations (blue line)
    resulting in an overshoot at $E=\chi$.  The orange line shows the
    result when the mesh is specifically tailored for material-based
    boundary conditions eliminating the overshoot at $v_{\const{x}}\approx
    0.5\,v_{th}$. Figure taken from a Ph.D.  dissertation
    \cite{Cagas2018}.}
  \label{fig:bc_R}
\end{figure}

This boundary condition can be precomputed and written as
automatically generated code with expanded matrix multiplications to
reduce computational cost significantly.  However, because it changes
based on the wall material and needs to be calculated for each cell,
it is stored as an external file.  As all the coefficients are
precomputed and the matrix multiplication is expanded, the actual
multiplication can be limited only to the non-zero terms, saving
computational time substantially.




Sheath simulations are performed in 1X1V using the dielectric boundary
conditions described in Sec.\thinspace\ref{sec:bc} and are compared to
sheath simulations that use an ideally absorbing wall.
\fgr{bc_bronold_diff} shows direct comparison (absolute difference in
the electron and ion distribution functions) of the simulation with
the dielectric boundary condition with the case that uses ideally
absorbing walls.  The solution is captured at $t\omega_{\const{pe}} =
500$ ($\omega_{\const{pe}}$ being the electron plasma frequency)
giving the simulations sufficient time to evolve from the same initial
conditions.  Note the periodic structure in the difference between the
electron distribution functions that results due to the absence of
Langmuir waves when using ideally absorbing walls.  More importantly,
note the higher electron density at the wall for the dielectric case
due to electron emission from the wall.  In the $v_{\const{x}}<0$ half
of the velocity domain, the acceleration of emitted electrons by the
sheath electric field is visible.  The ion distribution
(\fgr{bc_bronold_diff}b) shows that ions reach lower velocities in the
sheath for a given distance from the wall with the dielectric boundary
condition compared to the case with an absorbing wall.

\begin{figure}[!htb]
  \centering
  \includegraphics[width=0.8\linewidth]{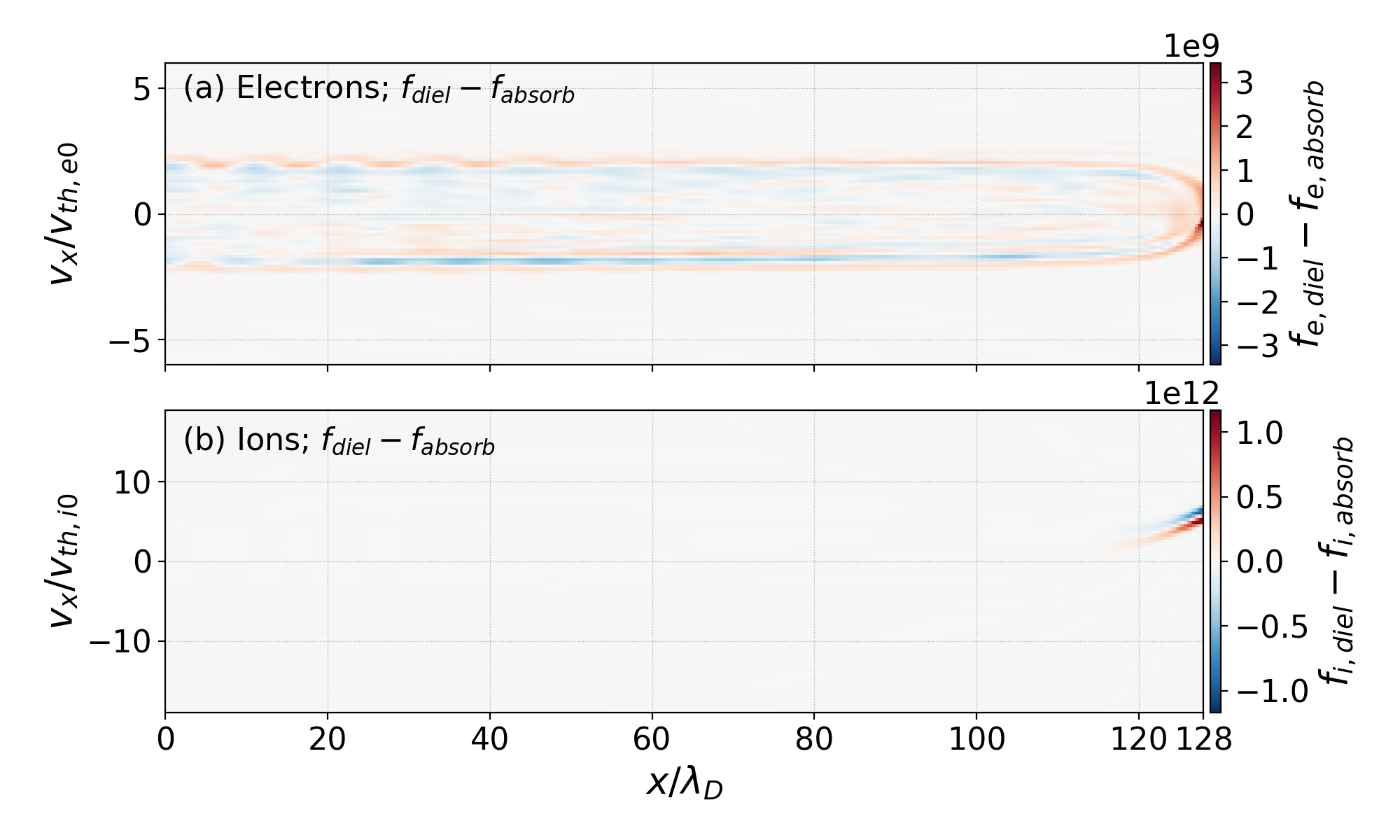}
  \caption{Direct comparison of electron and ion distribution
    functions from sheath simulations with absorbing and dielectric
    boundary conditions ($f_{diel}-f_{absorb}$).  The red color
    denotes regions with higher particle phase space density as is the
    case with the dielectric wall boundary condition.  The periodic
    structure is caused by the absence of Langmuir waves in the
    simulation that uses an ideally absorbing wall.  Data are
    presented at $t\omega_{\const{pe}}=500$. Figure taken from a Ph.D.
    dissertation \cite{Cagas2018}.}
  \label{fig:bc_bronold_diff}
\end{figure}

Plots of electron and ion densities, ion bulk velocity, electron and
ion temperatures, and electric fields are provided in
\fgr{bc_bronold_profiles}.  The simulation with the dielectric
boundary condition shows roughly twice the electron density of the
absorbing wall simulation in the region adjacent to the wall.
Returning electrons from the dielectric wall decrease the overall
outflow from the domain resulting in significantly smaller electric
field needed to equalize the electron and ion fluxes as noted in
\fgr{bc_bronold_profiles}(b).  Also note that the dielectric wall
reduces the potential difference between the wall and the plasma
compared to the absorbing wall.  The vertical solid black line in
\fgr{bc_bronold_profiles} marks the Bohm velocity crossing for both
cases which can be considered an approximation for the sheath edge.
Note that the differences between the solutions for the dielectric
boundary condition and the ideally absorbing boundary condition are
localized inside the sheath region.  An exception is noted through the
small differences in the presheath electric field caused by Langmuir
waves in the absorbing wall simulations.  As a result of the presheath
generally being unaffected by the dielectric wall, ions have the same
presheath acceleration profiles and reach the Bohm velocity at the
same distance from the wall for both cases.  The most significant
difference is in the electron temperature (\fgr{bc_bronold_profiles}d)
inside the sheath with lower gradients in thermal velocity for the
dielectric wall compared to the absorbing wall.

\begin{figure}[!htb]
  \centering
  \includegraphics[width=0.8\linewidth]{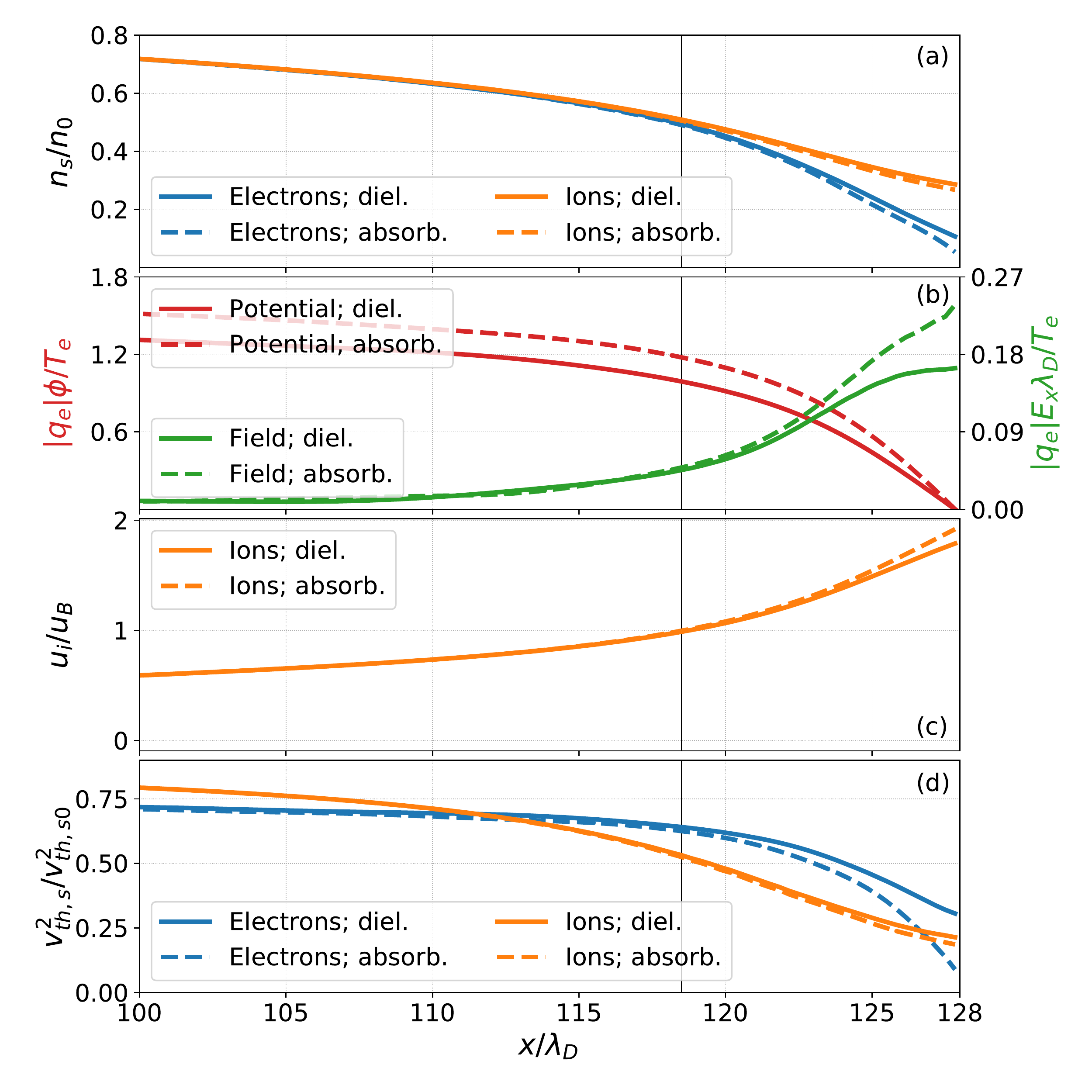}
  \caption{Comparison of profiles from sheath simulations with
    absorbing and dielectric wall boundary conditions. From top to
    bottom, panels show density (a), electric field and electrostatic
    potential (b), ion bulk velocity (c), and temperature (d).  In
    each of the panels, the solid line marks simulations with the
    dielectric wall boundary condition based on \eqr{bronold_R} while
    the dashed lines corresponds to simulations with an ideally
    absorbing wall.  Vertical dashed line marks the crossing of the
    Bohm velocity.  Data are presented at
    $t\omega_{\const{pe}}=500$. No collisions or ionization are
    included in these simulations. Figure adapted from a Ph.D.
    dissertation \cite{Cagas2018}.}
  \label{fig:bc_bronold_profiles}
\end{figure}

Higher moments of the distribution function are explored to understand
the significant differences in the electron temperature inside the
sheath between the dielectric and absorbing wall simulations.  The
1X1V simulation of \fgr{bc_bronold_profiles} produces a scalar value
for the third moment of the distribution instead of the full heat flux
tensor,
\begin{equation*}
  q_{\const{e}}(x) = \frac{1}{2}m_{\const{e}} \int_{-\infty}^\infty v_{\const{x}}^3 f_{\const{e}}(x,v_{\const{x}}) \, \dif v_{\const{x}}.
\end{equation*}
Normalized profile of $q_{\const{e}}$ in the region near the wall is
shown in \fgr{bc_bronold_qprofiles}a.  Due to the $v_{\const{x}}^3$
term, the third moment is particularly sensitive to oscillations of
the distribution function like the Langmuir waves discussed in
previous work \cite{Cagas2018}.  Therefore, the results in
\fgr{bc_bronold_qprofiles} are averaged over $\Delta
t\omega_{\const{pe}} = 1000$.

\fgr{bc_bronold_qprofiles}a shows that the heat flux to the wall is
higher for the case with the dielectric wall, which might seem to
contradict the lower temperature gradients shown in
\fgr{bc_bronold_profiles}d.  However, $q_{\const{e}}$ describes an
energy flux, i.e., it includes the local particle density which is
much higher for the case with the dielectric wall.  The quantity
plotted in \fgr{bc_bronold_qprofiles}a is normalized to the initial
number density in the center of the domain so the result is
dimensionless.  Alternatively, the third moment can be normalized to
the local number density, $q_{\const{e}}(x)/n_{\const{e}}(x)$.
\fgr{bc_bronold_qprofiles}b shows this quantity compared for the
dielectric and absorbing wall cases.  The lower flux in the dielectric
case is in agreement with the higher electron temperature and lower
electron temperature gradients inside the sheath
(\fgr{bc_bronold_profiles}d).

\begin{figure}[!htb]
  \centering
  \includegraphics[width=0.8\linewidth]{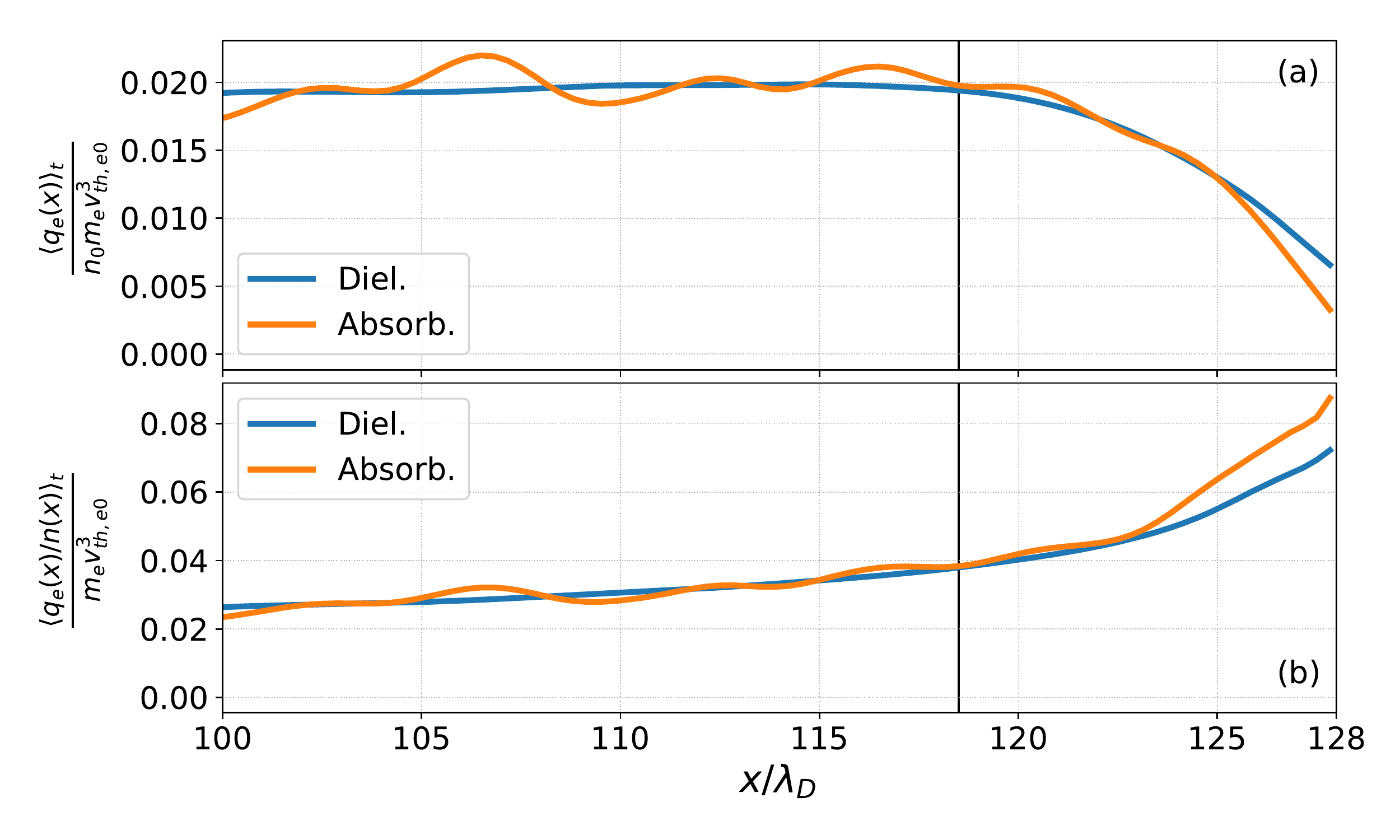}
  \caption{A comparison of heat flux profiles from the sheath
    simulations with absorbing and dielectric wall boundary
    conditions.  The top panel (a) shows the third moment of the
    distribution function, $q_{\const{e}}=\frac{1}{2}m_{\const{e}}\int
    v_{\const{x}}^3 f_e\dif v_{\const{x}}$, normalized to initial
    temperature and density, while the bottom panel (b) presents
    $q_{\const{e}}$ normalized to local density,
    $n_{\const{e}}(x)$. The profiles are averaged over the whole
    course of the simulation, $\Delta t\omega_{\const{pe}} = 1000$.}
  \label{fig:bc_bronold_qprofiles}
\end{figure}

\section{Conclusions}

A novel, self-consistent way to formulate boundary conditions through
general reflection functions for continuum kinetic simulations is
presented and its usage is demonstrated on simple specular reflection.
The same framework is then used for more complex electron surface
emission models---a phenomenological model \cite{Furman2002} and a
quantum mechanics based model \cite{Bronold2015}.

While the formulation of the boundary condition is general, it is
developed and presented using the discontinuous Galerkin method.  A
benchmark of a specular reflection boundary condition is implemented
to reproduce central symmetry of a plasma sheath simulation with
absorbing walls on both sides and no magnetic field.  After letting
the simulations evolve for $1000/\omega_{\const{pe}}$, maximum relative
differences between the full-domain case with two walls and the
half-domain case with reflecting boundary conditions are on the order
or $10^{-13}$, which corresponds to accumulated round-off error.

Finally, the quantum mechanics based model \cite{Bronold2015} is
self-consistently implemented for simulations of classical sheaths
with electron emission and is compared with an ideally absorbing wall.
The results show a significant impact on electron and ion profiles as
well as the electrostatic potential even for the simplest case of a
one-dimensional sheath in each of the configuration and velocity space
dimensions.  With the novel boundary condition, electron density at
the wall is doubled and electric field magnitude is roughly 40\% lower
in comparison to the case with ideally absorbing walls.  This work
presents the first description and implementation of a generalized
framework to incorporate high-fidelity electron emission and surface
physics boundary conditions into a continuum-kinetic plasma code.

\section*{Acknowledgements}

Authors are grateful for insights from conversations with James Juno
and other members of the \texttt{Gkeyll} team.  Simulations were
performed at the Advanced Research Computing center at Virginia Tech
(\url{http://www.arc.vt.edu}).  This research was supported by the Air
Force Office of Scientific Research under grant number
FA9550-15-1-0193.

\bibliographystyle{plainnat}
\bibliography{reference.bib}

\appendix
\section{Getting \texttt{Gkeyll} and reproducing the results}\label{app:gke}

To allow interested readers to reproduce our results and also use
\texttt{Gkeyll} for their applications, this Appendix provides
instructions to get the code (in both binary and source format). Full
installation instructions for \texttt{Gkeyll} are provided on the
\texttt{Gkeyll} website (\url{http://gkeyll.readthedocs.io}). The code
can be installed on Unix-like operating systems (including Mac OS and
Windows using the Windows Subsystem for Linux) either by installing
the pre-built binaries using the \textit{conda} package manager
(\url{https://www.anaconda.com}) or building the code via sources.
Input files for the simulations presented here will be made available
upon request.
\end{document}